\documentclass[journal]{IEEEtran}
\IEEEoverridecommandlockouts
\usepackage{setspace}
\usepackage{graphicx}
\usepackage{amssymb}
\usepackage{amsmath,amsfonts,amssymb,euscript,epsfig,psfrag,
enumerate,float,afterpage, subfigure}%

\newcommand{\expect}[1]{\mathbb{E}\left\{#1\right\}}
\newcommand{\defequiv}{\mbox{\raisebox{-.3ex}{$\overset{\vartriangle}{=}$}}}

\usepackage{cite}

\begin{document}

\title{Delay-Limited Cooperative Communication with Reliability Constraints in Wireless Networks}

%\author{Rahul Urgaonkar, Michael J. Neely 
%\\ University of Southern California, Los Angeles, CA 90089 
%\\ http://www-scf.usc.edu/$\sim$urgaonka
%$\vspace{-.4in}$
%\thanks{This material is supported in part  by one or more of
%the following: the DARPA IT-MANET program
%grant W911NF-07-0028, the NSF grant OCE 0520324,
%the NSF Career grant CCF-0747525.}}

\author{Rahul~Urgaonkar, Michael~J.~Neely \\ 
%\{urgaonka, mjneely \}@usc.edu\\
%University of Southern California, Los Angeles, CA 90089\\ 
%http://www-scf.usc.edu/$\sim$urgaonka\\
%$\vspace{-.4in}$
\IEEEcompsocitemizethanks{\IEEEcompsocthanksitem This work was presented in part at the IEEE INFOCOM conference,
Rio de Janeiro, Brazil, April 2009.
\IEEEcompsocthanksitem Rahul Urgaonkar and Michael J. Neely are with the Department
of Electrical Engineering, University of Southern California, Los Angeles, CA
90089. Web: http://www-scf.usc.edu/$\sim$urgaonka
\IEEEcompsocthanksitem This material is supported in part  by one or more of
the following: the DARPA IT-MANET program
grant W911NF-07-0028, the NSF grant OCE 0520324,
the NSF Career grant CCF-0747525.}}

\maketitle
\begin{abstract}
We investigate optimal resource allocation for delay-limited
cooperative communication in time varying wireless networks.
Motivated by real-time applications that have stringent delay
constraints, we develop a dynamic cooperation strategy that makes
optimal use of network resources to achieve a target outage
probability (reliability) for each user subject to average power
constraints. Using the technique of Lyapunov optimization, we first
present a general framework to solve this problem and then derive
quasi-closed form solutions for several cooperative protocols
proposed in the literature. Unlike earlier works, our scheme
does not require prior knowledge of the statistical description of
the packet arrival, channel state and node mobility processes and
can be implemented in an online fashion.
%Both scenarios where channel state
%information is available at the transmitter and when only the
%statistics are known are considered.
\end{abstract}
%$\vspace{-.5in}$

\section{Introduction}
\label{section:intro}

There is growing interest in the idea of utilizing cooperative
communication \cite{NOW_survey, laneman_survey, Laneman1, Laneman2, Sendonaris1, Sendonaris2} to
improve the performance of wireless networks with time varying
channels. The motivation comes from the work on MIMO systems
\cite{tsebook} which shows that employing multiple antennas on a
wireless node can offer substantial benefits. However, this may be
infeasible in small-sized devices due to space limitations.
Cooperative communication has been proposed as a means to achieve
the benefits of traditional MIMO systems using \emph{distributed
single antenna} nodes. Much recent work in this area promises
significant gains in several metrics of interest (such as diversity \cite{Laneman1}\cite{Laneman2}, capacity
 \cite{Sendonaris1, Sendonaris2, gastpar1, kramer1, host-madsen}, energy efficiency \cite{alouini, wanjen_survey}, etc.) over
conventional methods. We refer the interested reader to a recent
comprehensive survey \cite{NOW_survey} and its references.

%In multi-user wireless networks with time varying fading channels,
%many cooperative strategies have been shown to outperform
%non-cooperative ones.

\begin{figure}
\centering
\includegraphics[width=9cm,  angle=0]{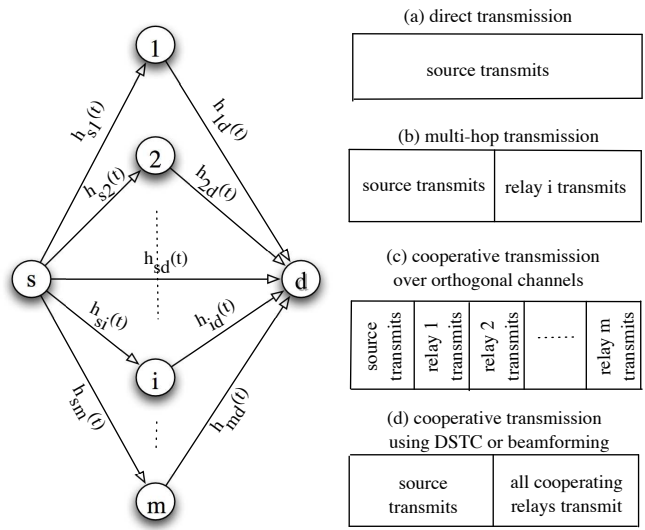}
\caption{Example $2$-hop network with source, destination and
relays. The time slot structures for different transmission
strategies are also shown. Due to the half-duplex constraint,
cooperative protocols need to operate in two phases. Hence, there is an
inherent loss in the multiplexing gain under any such cooperative
transmission strategy over direct transmission.}\label{fig:one}
\end{figure}

 The main idea behind cooperative communication can be
understood by considering a simple $2$-hop network consisting of a
source $s$, its destination $d$ and a set of $m$ relay nodes  as
shown in Fig. \ref{fig:one}. Suppose $s$ has a packet to send to $d$
in timeslot $t$. The channel gains for all links in this network are
shown in the figure. In direct
communication, $s$ uses the full slot to transmit its packet to $d$
over link $s-d$ as shown in Fig. \ref{fig:one}(a). In conventional multi-hop relaying, $s$ uses the
first half of the slot to transmit its packet to a particular relay node $i$
over link $s-i$ as shown in Fig. \ref{fig:one}(b). If $i$ can successfully decode the packet, it
re-encodes and transmits it to $d$ in the second half of the slot
over link $i-d$. In both scenarios, to ensure reliable
communication, the source and/or the relay must transmit at high
power levels when the channel quality of any of the links involved
is poor. However, note that due to the broadcast nature of wireless
transmissions, other relay nodes may receive the signal from the
transmission by $s$ and can cooperatively relay it to $d$. The
destination now receives multiple copies/signals and can use all of
them jointly to decode the packet. Since these signals have been
transmitted over independent paths, the probability that all of them
have poor quality is significantly smaller. Cooperative
communication protocols take advantage of this \emph{spatial
diversity gain} by making use of multiple relays for cooperative
transmissions to increase reliability and/or reduce energy costs.
This is different from traditional multi-hop relaying in which only
one node is responsible for forwarding at any time and in which the
destination does not use multiple signals to decode a packet.

Because of the half-duplex nature of wireless devices, a relay node
cannot send and receive on the same channel simultaneously.
Therefore, such cooperative communication protocols typically
operate over a two phase slot structure as shown in Figs.
\ref{fig:one}(c) and \ref{fig:one}(d). In the first phase, $s$ transmits its packet to the
set of relay nodes. In the second phase, a subset of these relays
transmit their signals to $d$. Note that the destination may receive
the source signal from the first phase as well. At the end of the
second phase, the destination appropriately combines all of these
received signals to decode the packet. The exact slot structure as
well as the signals transmitted by the relays depend on the
cooperative protocol being used.\footnote{We consider several
protocol examples in Sec. \ref{section:2stage}} For example, Fig. \ref{fig:one}(c) shows the
slot structure under a cooperative scheme that transmits over orthogonal channels. Specifically, the
time slot is divided into $m+1$ equal mini-slots. In phase one, the source transmits its packet 
in the first mini-slot. In the second phase, the relays transmit one after the other in their
own mini-slots. Fig. \ref{fig:one}(d) shows the slot structure under a cooperative scheme in which the cooperating
relays use distributed space-time codes (DSTC) or a beamforming technique to transmit simultaneously in the second phase. 
 It should be noted that due to this half-duplex constraint, 
there is an inherent loss in the multiplexing gain
under any such cooperative transmission strategy over direct transmission.
Therefore, it is important to develop algorithms that cooperate
opportunistically.

In this work, we consider a mobile ad-hoc network with 
\emph{delay-limited} traffic and cooperative communication. Many
real-time applications (e.g., voice) have stringent delay constraints
and fixed rate requirements. In slow fading environments
(where decoding delay is of the order of the channel coherence
time), it may not be possible to meet these delay constraints for
every packet. However, these applications can often tolerate a
certain fraction of lost packets or outages. A variety of techniques
are used to combat fading and meet this target outage probability
(including exploiting diversity, channel coding, ARQ, power control,
etc.). Cooperative communication is a particularly attractive
technique to improve reliability in such delay-limited scenarios
since it can offer significant spatial diversity gains in addition
to these techniques.

%\footnote{ Relevant to such scenarios are the information theoretic
%notions of delay-limited capacity \cite{tse2} and minimum outage
%probability \cite{caire1} in a multi-access fading channel}

Much prior work on cooperative communication considers physical
layer resource allocation for a static network, particularly in the
case of a single source.  
Objectives such as minimizing sum power,
minimizing outage probability, meeting a target SNR constraint,
etc., are treated in this context \cite{host-madsen, alouini, wanjen_survey, Maric1,
Maric2, Adve1, gunduz, minchen}. We draw on this work in the development of
\emph{dynamic} resource allocation in a stochastic network with
fading channels, node mobility, and random packet arrivals, where
\emph{opportunistic cooperation decisions} are required. Dynamic
cooperation was also considered in the prior work \cite{yeh} which 
investigates throughput optimality and queue stability in a multi-user network with static channels
and randomly arriving traffic using the framework of
Lyapunov drift. Our formulation is different and does not involve
issues of queue stability.  Rather, we consider a delay-limited
scenario where each packet must either be transmitted in one slot,
or dropped.  This is similar to the concept of \emph{delay-limited
capacity} \cite{tse2}. Also related to such scenarios is the notion of
\emph{minimum outage probability} \cite{caire1}. 
These quantities are also investigated in the recent work \cite{gunduz} 
that considers a $3$ node static network with Rayleigh fading and shows that 
opportunistic cooperation significantly improves the delay-limited capacity.

%We focus on the single-source setting with
%mobility case in this paper and develop an algorithm that optimizes a weighted sum of
%reliability and power expenditure subject to individual reliability
%and average power constraints at the source and all relays. We then
%extend this to a multiple-source setting in \cite{urgaonkar-coop}.

In this work, we use techniques of both Lyapunov drift and Lyapunov optimization
\cite{neely-NOW} to develop a control algorithm that takes dynamic decisions for each new slot. 
Different from most work that applies this theory,
our solution involves a $2$-stage stochastic shortest path problem
due to the cooperative relaying structure.  This problem is
non-convex and combinatorial in nature  and does not admit closed form solutions in general.
However, under several important and well known classes of physical layer cooperation
models, we develop techniques for reducing the problem exactly to an
$m$-stage set of convex programs.  The convex programs themselves
are shown to have quasi-closed form solutions and can be computed in
real time for each slot, often involving simple water-filling strategies that also
arise in related static optimization problems.

%\emph{An information theoretic definition and characterization of
%delay-limited capacity in a multi-access fading channel is presented
%in \cite{tse2}. Also relevant to such scenarios is the notion of
%minimum outage probability \cite{caire1}. 
%In this work, we consider
%the related problem of optimal resource allocation to meet a target
%outage probability requirement. Note that these works treat
%static networks and require knowledge of channel fade distributions..}

 %as follows. \cite{tse2} looks at delay-limited capacity under both
%long term and peak power constraints.
%
%\cite{caire1} considers the problem of minimizing the outage
%probability under both short term and long term power constraints.
%The relationship between delay-limited capacity and minimum outage
%probability is that the former is the maximum rate at which the
%ater is zero.

%\cite{jindal} considers DMT...

%For example, its shown that can get non-zero delay-limited capacity
%with cooperative communication (Gunduz) in contrast to the direct
%transmission case?

%\section{Basic Model and Control Objective}
%\label{section:basic}

\section{Basic Network Model}
\label{section:basic}

We consider a mobile ad-hoc network with delay-limited communication
over time varying fading channels. The network contains a set
$\mathcal{N}$ of nodes, all potentially mobile. 
All nodes are assumed to be within range of each other, and any node
pair can communicate either through direct transmission or through a
$2$-phase cooperative transmission that makes use of other nodes as
relays. The system operates in slotted time  and the channel coefficient between
nodes $i$ and $j$ in slot $t$ is denoted by $h_{ij}(t)$. We assume a
block fading model \cite{tsebook} for the channel coefficients  so that their value
remains fixed during a slot and changes from one slot to the other
according to the distribution of the underlying fading and mobility
processes.

For simplicity, we assume that the set
$\mathcal{N}$ contains a single source node $s$ and its destination
node $d$ and that all other nodes act simply as cooperative relays.
This is similar to the single-source assumption treated in
\cite{Maric1, Maric2, gunduz, minchen, Adve1} for static networks. We derive
a dynamic cooperation strategy for this single source problem in
Sec. \ref{section:CNC} that optimizes a weighted sum of
reliability and power expenditure subject to individual reliability
and average power constraints at the source and at all relays. 
 This highlights the decisions
involved from the perspective of a source node, and these decisions
and the resulting solution structure are similar to the multi-source
scenario operating under an orthogonal medium access scheme (such as TDMA or FDMA) 
studied later in Sec. \ref{section:extensions}.
In the following, we denote the set of relay nodes by $\mathcal{R}$ and the set $\{s\}
\cup \mathcal{R}$ by $\mathcal{\widehat{R}}$. All nodes $i \in \mathcal{\widehat{R}}$ have both long
term average and instantaneous peak power constraints given by
$P_i^{avg}$ and $P_i^{max}$ respectively.

% under a
%natural orthogonal multi-access structure,  such as TDMA or FDMA.
%We then extend this to a multiple-source setting in \cite{urgaonkar-coop}.
%This solution is extended in \cite{urgaonkar-coop} 
%to treat the multi-source problem. 

We consider two models for the availability of the
channel state information (CSI). The first is the 
\emph{known channels, unknown statistics} model.
Under this model, we assume that 
the channel gains between the source node and its relay set
and destination as well as the channel gains between the relays and the
destination are known every slot. These could be obtained by sending pilot
signals and via feedback. This model has also been considered in prior works \cite{Maric1, Maric2, gunduz, minchen} on power allocation
in static networks where, in addition to the current channel gains, a knowledge of the
distribution governing the fading process is assumed.
In our work, under this \emph{known channels, unknown statistics} model, we do not assume any knowledge
of the distributions governing the evolution of the channel states, mobility processes, or traffic. 
Thus, our algorithm and its optimality properties hold for
a very general class of channel and mobility models that satisfy certain ergodicity requirements (to be made precise later).
We note that the channel gain could represent just the amplitude of the channel coefficient if an orthogonal
cooperative scheme is being used. However, in case of cooperative schemes such as beamforming, this could
represent the complete description of the fading coefficient that includes the phase information.

The second model we consider is the \emph{unknown channels, known statistics} model.
In this case, we assume that the current set of
potential relay nodes is known on each slot $t$, but the exact channel
realizations  between the source and these relays, and the relays and
the destination, are unknown.  Rather, we assume only that the \emph{statistics}
of the fading coefficients are known between the source and current relays,
and the current relays and destination. However, we still do not require
knowledge of the distributions governing the arriving traffic or
the mobility pattern (which affects the set of relays we will see in
future slots). This is in contrast to prior works that have considered
resource allocation in the presence of partial CSI only for static networks.

%In the second model for the availability of the channel state information, the current channel gains are not known. 
%Rather, we assume that the statistics of the fading coefficients are known. 
%However, we do not assume any knowledge of the mobility patterns.
%This is in contrast to prior works that have considered resource allocation in the presence of partial CSI only for
%static networks. 

For both models, we use $\mathcal{T}(t)$ to represent the collection of all channel
state information known on slot $t$.  For the known channels, unknown
statistics model, $\mathcal{T}(t)$ represents the collection of channel coefficients
$h_{ij}(t)$ between the source and relays and relays and destination.
For the unknown channels, known statistics model, $\mathcal{T}(t)$ represents
the set of all nodes that are available on slot $t$ for relaying and the distribution of the
fading coefficients. We assume that $\mathcal{T}(t)$ lies in a space of finite but arbitrarily large size
and evolves according to an ergodic process with a well defined
steady state distribution.  This variation in channel state
information affects the reliability and power expenditure associated
with the direct and cooperative transmission modes that are discussed
in Sec. \ref{section:options}.

%The collection of all such channel state information
%in slot $t$ is represented by $\mathcal{T}(t)$. We assume that this
%lies in a space of finite but arbitrarily large size and evolves
%according to an ergodic process with a well defined steady state
%distribution. This variation in channel quality  affects the reliability
%and power expenditure associated with the direct and the cooperative
%transmission options that are discussed next. 

%To the best of our knowledge, there are no prior works that treat this model for mobile networks... (laneman?)

\subsection{Example of Channel State Information Models}
\label{section:example}

As an example of these models, suppose the nodes move in a cell-partitioned
network according to a Markovian random walk (see also Fig. \ref{fig:cell} in Sec. \ref{section:sim} on Simulations). 
Each slot, a node may decide to stay in its current cell or move
to an adjacent cell according to the probability distribution governing the
random walk. Suppose that each slot, the set of potential relays consists only
of nodes in either the same or an adjacent cell of the source.  
Suppose channel gains between nodes in the same cell are distributed according to a
Rayleigh fading model
%\footnote{Here the coefficients are modeled as zero-mean, independent, circularly
%symmetric complex Gaussian random variables
% and the magnitudes $|h_{ij}(t)|$ are Rayleigh distributed 
%with $\expect{|h_{ij}(t)|^2} = 1/{d_{ij}^\alpha (t)}$ (where $d_{ij}(t)$ denotes the distance
%between node $i$ and $j$ is slot $t$ and $\alpha$ denotes the path loss exponent).} 
with a particular mean and variance, while
gains for nodes in adjacent cells are Rayleigh with a different mean and variance.
Under the {known channels, unknown statistics} model, the $\mathcal{T}(t)$ information is the set of current
gains $h_{ij}(t)$, and the Rayleigh distribution is not needed.  Under the
{unknown channels, known statistics} model, the $\mathcal{T}(t)$ information is the set of
nodes currently in the same and adjacent cells of the source, and we assume
we know that the fading distribution is Rayleigh, and we know the corresponding
means and variances. However, neither model requires knowledge of
the mobility model or the traffic rates.

%As an example of these models, suppose the nodes move in a cell-partitioned network according to
%a Markovian random walk. Each slot, a node may decide to stay in its current cell 
%or move to a neighboring cell according to the probability distribution governing the random walk.
%Suppose in each slot, the channel gains evolve according to a Rayleigh fading
%model. 
%in a Rayleigh channel fading model, 
%Then, under the first model, we assume that the current channel gains $h_{ij}(t)$ are known
%every slot. However, no knowledge of the channel fading model or mobility patterns is assumed. Under the second model,
%we assume that the current distribution of $h_{ij}(t)$ is known.

%and every slot the channel
%conditions between nodes can change. 

%In the basic model, we assume that the set $\mathcal{N}$ contains a
%single source node $s$, its destination node $d(s)$ and that all
%other nodes act simply as cooperative relays.

\subsection{Control Options}
\label{section:options}

 Suppose the slot size is normalized to integer slots $t \in \{0, 1, 2, \ldots, \}$. In each
slot, the source $s$ receives new packets for its destination $d$
according to an i.i.d. Bernoulli process $A_s(t)$ of rate
$\lambda_s$. Each packet is assumed to be $R$ bits long and has a
\emph{strict} delay constraint of $1$ slot. Thus, a packet not
served within $1$ slot of its arrival is dropped. Further, packets
that are not successfully received by their destinations due to
channel errors are not retransmitted. The source node has a minimum
time-average reliability requirement specified by a fraction
$\rho_s$ which denotes the fraction of packets that were transmitted
successfully. In any slot $t$, if source $s$
has a new packet for transmission, it can use one of the following
%control options 
transmission modes (Fig. \ref{fig:one}):
\begin{enumerate}
\item Transmit directly to $d$ using the
full slot
\item Transmit to $d$ using traditional relaying
over two hops
\item Transmit cooperatively with the set $\mathcal{R}$ of relay
nodes using the two phase slot structure
% as described before
\item Stay idle  (so that the packet gets dropped)
\end{enumerate}

We consider all of these transmission modes because, depending on the current
channel conditions and energy costs in slot $t$, it might be better to choose
one over the other.  For example, due to the half-duplex constraint,
direct transmission using the full slot might be preferable to
cooperative transmission over two phases on slots when the source-destination
link quality is good. Note that this is similar to the much studied framework of opportunistic transmission
scheduling in time varying channels.
Further, even in the special case of static channels, 
the optimal strategy may involve a mixture of these modes of operation to meet
the target reliability and average power constraints.

%For example, in the case of a single fading channel, the objective
%of minimizing the outage probability under a long term power
%constraints is considered in \cite{caire1} where a threshold rule is
%derived for transmission....
%the optimal decision is to not transmit and take an outage.

%In fact, sometimes it might be best not to
%transmit at all (and take an outage) because of bad channels and the
%resulting high cost of transmission.

Let $\mathcal{I}^\eta(t)$ denote the collective control action in slot $t$ under some policy $\eta$ that includes 
the choice of 
the transmission mode at the source,
power allocations for the source and all relevant relays, and any additional
physical layer choices such as modulation and coding.
Specifically, we have:
\begin{align*}
%\mathcal{I}^\eta(t) = [\textrm{mode choice}, \overrightarrow{P}^\eta(t), \textrm{other PHY layer choices}]
\mathcal{I}^\eta(t) = [\textrm{mode choice}, \textbf{\emph{P}}^\eta(t), \textrm{other PHY layer choices}]
\end{align*}
where the mode choice refers to one of the $4$ transmission modes for
the source, and where 
%$\overrightarrow{P}^\eta(t)$ 
$\textbf{\emph{P}}^\eta(t)$ 
is the collection of coefficients $P_i^\eta(t)$ representing power
allocations for each node $i \in \mathcal{\widehat{R}}$. Note that $P_i^\eta(t) = 0$ for all $i$ under
transmission mode $4$ (idle).  If the source $s$ chooses mode $1$, we have $P_i(t) = 0$
for all relay nodes $i \in \mathcal{R}$, whereas if $s$ chooses mode $2$, we have $P_i(t) > 0$ for
at most one relay $i \in \mathcal{R}$.  
Note that under any feasible policy $\eta$, $P_i^\eta(t)$ must satisfy the instantaneous peak power constraint every slot 
for all $i$. Also note that under the cooperative transmission option, the power allocation for the source 
node and the relays corresponds to the first and second phase respectively. Thus, the source is active in the first phase while the relays 
are active in the second phase. We denote the set of all valid power allocations by $\mathcal{P}$ and
define $\mathcal{C}$ as the set of all valid control actions:
\begin{align*}
\mathcal{C} = \{1, 2, 3, 4\} \times \{ \mathcal{P} \} \times \{ \textrm{other PHY layer choices} \}
\end{align*}

The success/failure outcome of the control action is represented by an indicator random variable 
$\Phi_s(\mathcal{I}^\eta(t), \mathcal{T}(t))$  that
depends on the current control action and channel state. 
Successful transmission of a packet is usually a complicated function of the
transmission mode chosen, the associated power allocations and channel
states, as well as physical layer details like modulation,
coding/decoding scheme, etc. In this work, the particular
physical layer actions are included in the $\mathcal{I}^\eta(t)$ decision variable. 
Specifically, given a control action $\mathcal{I}^\eta(t)$ and a
channel state $\mathcal{T}(t)$, the outcome 
is defined as follows:
%\begin{displaymath}
\begin{align}
\Phi_s(\mathcal{I}^\eta(t),
\mathcal{T}(t)) \defequiv \left\{ \begin{array}{lll} 1 & \textrm{if a packet transmitted by $s$ in slot} \\
 & \textrm{$t$ is successfully received by $d$} \\
0 & \textrm{else}
\end{array} \right.
\label{eq:phi}
\end{align}
%\end{displaymath}

Note that $\Phi_s(\mathcal{I}^\eta(t), \mathcal{T}(t))$ is a random
variable, and its conditional expectation given $(\mathcal{I}^\eta(t), \mathcal{T}(t))$ is equal to the success probability
under the given physical layer channel model.  
Use of this abstract indicator variable allows a unified treatment that
can include a variety of physical layer models.  Under the known channels,
unknown statistics model (where $\mathcal{T}(t)$ includes the full channel realizations
between source and relays and relays and destination on slot $t$), 
$\Phi_s(\mathcal{I}^\eta(t), \mathcal{T}(t))$ can be a determinisitic $0/1$ function based on the known channel state
and control action. Specific examples for this model are considered in Sec. \ref{section:2stage}.  
Under the unknown channels, known statistics model (where $\mathcal{T}(t)$ 
represents only the set of current possible relays and the fading statistics), 
we assume we know the value of $Pr[\Phi_s(\mathcal{I}^\eta(t), \mathcal{T}(t))=1]$
under each possible control action $\mathcal{I}^\eta(t)$.  This model is considered in Sec. \ref{section:stats}.
Under both models, we assume that explicit ACK/NACK information is received at the end
of each slot, so that the source knows the value of $\Phi_s(\mathcal{I}^\eta(t), \mathcal{T}(t))$.
 For notational convenience, in the rest of the paper, we use $\Phi_s^\eta(t)$ instead of
$\Phi_s(\mathcal{I}^\eta(t), \mathcal{T}(t))$ noting that the dependence on $(\mathcal{I}^\eta(t), \mathcal{T}(t))$ is implicit.

\subsection{Discussion of Basic Model}
\label{section:discuss}

The basic model described above extends prior work on $2$-phase
cooperation in static networks to a mobile environment, and treats
the important example scenario where a team of nodes move in a tight cluster but
with possible variation in the relative locations of nodes within
the cluster. We note that our model and results are applicable to the 
special case of a static network 
%(with no channel variation) 
as well. Another example scenario captured by our model is an OFDMA-based cellular
network with multiple users that have both inter-cell and intra-cell
mobility. In each slot, a set of transmitters is determined in each
orthogonal channel (for example, based on a predetermined TDMA
schedule, or dynamically chosen by the base station). The remaining nodes
can potentially act as cooperative relays in that slot. 
%In such networks, the applications have tight delay constraints...

The basic model treats scenarios in which a source node
can transmit to its destination, possibly with the help of multiple
relay nodes, in $2$ stages. While this is a simplifying assumption,
the framework developed here can be applied to more general scenarios in which, in a single slot, cooperative relaying
over $K$ stages is performed (for some $K > 2$) using
multi-hop cooperative techniques (e.g., \cite{scaglione, shashi}).

%possibly usign the multi-hop cooperative techniques used for queue stability problems in [Yeh]

%In a more general model, a relay could simultaneously cooperate with
%multiple transmitters if they are transmitting on orthogonal
%channels. Further, a transmitting node may simultaneously act as a
%relay for other nodes transmitting on orthogonal channels. We do not
%consider these scenarios in this work.

\section{Control Objective}
\label{section:objective}

Let $\alpha_s$ and $\beta_i$ for $i \in \mathcal{\widehat{R}}$ be a collection of non-negative
weights. Then our objective is to design a policy $\eta$ that solves
the following \emph{stochastic optimization problem}:

\begin{align}
\textrm{Maximize:} \qquad & \alpha_s \bar{r}^\eta_s - \sum_{i\in \mathcal{\widehat{R}}} \beta_i \bar{e}^\eta_i\nonumber \\
\textrm{Subject to:} \qquad &  \bar{r}^\eta_s \geq \rho_s \lambda_s \nonumber \\
&  \bar{e}^\eta_i \leq P_i^{avg} \; \forall \; i \in \mathcal{\widehat{R}} \nonumber \\
&  0 \leq P_i^\eta(t) \leq P_i^{max} \; \forall \; i \in \mathcal{\widehat{R}}, \; \forall t \nonumber \\
& \mathcal{I}^\eta(t) \in \mathcal{C} \; \forall t 
\label{eq:obj1}
\end{align}
where $\bar{r}_s^\eta$ is the time average reliability for source $s$
under policy $\eta$ and is defined as:
\begin{align}
&\bar{r}_s^\eta \defequiv \lim_{t\rightarrow\infty}
\frac{1}{t}\sum_{\tau=0}^{t-1}\expect{\Phi_s^\eta(\tau)}
\label{eq:ri}
\end{align}

and $\bar{e}_i^\eta$ is the time average power usage of node $i$ under $\eta$:

\begin{align}
&\bar{e}_i^\eta \defequiv \lim_{t\rightarrow\infty}
\frac{1}{t}\sum_{\tau=0}^{t-1}\expect{P_i^\eta(\tau)} \label{eq:ei}
\end{align}

Here, the expectation is with respect to the possibly randomized
control actions that policy $\eta$ might take. The $\alpha_s$ and
$\beta_i$ weights allow us to consider several different objectives.
For example, setting $\alpha_s = 0$ and $\beta_i = 1$ for all $i$
reduces (\ref{eq:obj1}) to the problem of minimizing the average sum
power expenditure subject to minimum reliability and average power
constraints. This objective can be important in the multiple source
scenario when the resources of the relays must be shared across many
users. Setting all of these weights to $0$ reduces (\ref{eq:obj1})
to a feasibility problem where the objective is to provide minimum
reliability guarantees subject to average power constraints.
%Or is it with respect to the random channel state distribution?

Problem (\ref{eq:obj1}) is similar to the general stochastic utility
maximization problem presented in \cite{neely-NOW}. Suppose
(\ref{eq:obj1}) is feasible and let $r^*_s$ and $e^*_i \; \forall i
\in \mathcal{\widehat{R}}$ denote the optimal value of the objective
function, potentially achieved by some arbitrary policy. Using the
techniques developed in \cite{neely-NOW,neely-energy}, it can be
shown that it is sufficient to consider only the class of
stationary, randomized policies that take control decisions purely
as a (possibly random) function of the channel state
$\mathcal{T}(t)$ every slot to solve (\ref{eq:obj1}).
However, computing the optimal stationary, randomized policy explicitly
can be challenging and often impractical as it requires knowledge of arrival distributions, channel
probabilities and mobility patterns in advance.  Further, as pointed out earlier, even in the
special case of a static channel, the optimal strategy may involve a mixture of
direct transmission, multi-hop, and cooperative modes of operation, and the relaying modes
must select different relay sets over time to achieve the optimal time average mixture.

However, the technique of Lyapunov optimization \cite{neely-NOW}
can be used to construct an alternate dynamic policy 
that overcomes these challenges and is provably optimal.
Unlike the stationary, randomized policy, this policy does not need to be computed beforehand
and can be implemented in an online fashion. 
%without a knowledge of the statistics of the arrivals, mobility patterns, etc.
 In the known channels model, it does not need a-priori
statistics of the traffic, channels, or mobility.  In the unknown channels
model, it does not need a-priori statistics of the traffic or mobility.
We present this policy in the next section.

\section{Optimal Control Algorithm}
\label{section:CNC}

In this section, we present a dynamic control algorithm that achieves 
the optimal solution $r^*_s$ and $e^*_i \; \forall
i \in \mathcal{\widehat{R}}$ to the stochastic optimization problem 
presented earlier. This algorithm is similar in spirit to the
backpressure algorithms proposed in \cite{neely-NOW, neely-energy}
for problems of throughput and energy optimal networking in time
varying wireless ad-hoc networks.

The algorithm makes use of a ``reliability queue'' $Z_s(t)$
for source $s$. Specifically, let $Z_s(t)$ be a value that is initialized to zero (so that $Z_s(0) = 0$),
and that is updated at the end of every slot $t$ according to the following equation:
\begin{align}
Z_s(t+1) = \max[Z_s(t)- \Phi_s(t),0] + \rho_s A_s(t)\label{eq:p1u1}
\end{align}
where $A_s(t)$ is the number of arrivals to source $s$ on slot $t$ (being either $0$ or $1$),
and $\Phi_s(t)$ is $1$ if and only if a packet that arrived was successfully delivered
(recall that ACK/NACK information gives the value of $\Phi_s(t)$ at the end of
every slot $t$). 
Additionally, it also uses the following virtual power queues
$\forall i \in \mathcal{\widehat{R}}$:
\begin{align}
X_i(t+1)=\max[X_i(t) - P_i^{avg},0] +  P_i(t) \label{eq:p1x1}
\end{align}
All these queues are also initialized to $0$ and updated at the end of every slot $t$ 
according to the equation above. We note that these queues are virtual in that they do not represent
any real backlog of data packets.  Rather, they facilitate the
control algorithm in achieving the time average reliability and
energy constraints of (\ref{eq:obj1}) as follows. If a policy $\eta$
stabilizes (\ref{eq:p1u1}), then we must have that its service rate
is no smaller than the input rate, i.e.,
\begin{align*}
\bar{r}^\eta_s = \lim_{t\rightarrow\infty}
\frac{1}{t}\sum_{\tau=0}^{t-1}\expect{\Phi^\eta_s(\tau)} \geq
\lim_{t\rightarrow\infty}\frac{1}{t}\sum_{\tau=0}^{t-1}\expect{\rho_s
A_s(\tau)} = \rho_s \lambda_s
\end{align*}

Similarly, stabilizing (\ref{eq:p1x1}) yields the following:
\begin{align*}
\bar{e}_i^\eta = \lim_{t\rightarrow\infty}
\frac{1}{t}\sum_{\tau=0}^{t-1}\expect{P_i^\eta(\tau)} \leq P_i^{avg}
\end{align*}
where we have used definitions (\ref{eq:ri}), (\ref{eq:ei}).
%\emph{Note:By ergodicity, the time average expected same as....}
This technique of turning time-average constraints into queueing
stability problems was first used in \cite{neely-energy}. 

To stabilize these virtual queues and optimize the objective function
in (\ref{eq:obj1}), the algorithm operates as follows. 
Let $\textbf{\emph{Q}}(t) = (Z_s(t), X_i(t)) \; \forall i \in \mathcal{\widehat{R}}$ denote the collection of these queues in timeslot $t$. 
Every slot $t$, given $\textbf{\emph{Q}}(t)$ and the current channel state $\mathcal{T}(t)$, it chooses a control action $\mathcal{I}^*(t)$ that
minimizes the following stochastic metric (for a given control parameter $V \geq 0$):
\begin{align}
\textrm{Minimize:} \qquad & (X_s(t) +V\beta_s)\expect{P_s(t)|\textbf{\emph{Q}}(t), \mathcal{T}(t)} + \nonumber \\
		& \sum_{i \in \mathcal{R}}(X_i(t)+V\beta_i)\expect{P_i(t)|\textbf{\emph{Q}}(t), \mathcal{T}(t)} - \nonumber \\
		& (Z_s(t)+V \alpha_s)\expect{\Phi_s(t)|\textbf{\emph{Q}}(t), \mathcal{T}(t)} \nonumber\\
\textrm{Subject to:} \qquad & 0 \leq  P_i(t) \leq P_i^{max} \; \forall i \in \mathcal{\widehat{R}} \nonumber \\
			& \mathcal{I}(t) \in \mathcal{C}
 \label{eq:p1ssp3}
\end{align}

After implementing $\mathcal{I}^*(t)$ and observing the outcome, the virtual queues are updated using (\ref{eq:p1u1}), (\ref{eq:p1x1}).
Recall that there are no actual queues in the system.  Our algorithm enforces
a strict $1$-slot delay constraint so that $\Phi_s(t) = 0$ if the packet is not successfully
delivered after $1$ slot.  The virtual queues $X_i(t), Z_s(t)$ are maintained only in
software and act as known weights in the optimization (\ref{eq:p1ssp3}) that guide decisions
towards achieving our time average power and reliability goals.  The control
action $\mathcal{I}^*(t$) that optimizes (\ref{eq:p1ssp3}) affects the powers $P_i(t)$ allocated and the
$\Phi_s(t)$ value according to (\ref{eq:phi}).

The above optimization is a $2$-stage \textit{stochastic shortest
path} problem \cite{bertsekas} where the two stages correspond to
the two phases of the underlying cooperative protocol. Specifically,
when $s$ decides to use the option of transmitting cooperatively,
the cost incurred in the first stage is given by the first term
$(X_s(t) +V\beta_s)\expect{P_s(t)|\textbf{\emph{Q}}(t), \mathcal{T}(t)}$. The cost incurred during the
second stage is given by $\sum_{i \in \mathcal{R}}(X_i(t)+ V\beta_i)\expect{P_i(t)|\textbf{\emph{Q}}(t), \mathcal{T}(t)}$ and at the end of this
stage, we get a reward of $(Z_s(t)+V \alpha_s)\expect{\Phi_s(t)|\textbf{\emph{Q}}(t), \mathcal{T}(t)}$.
The transmission outcome $\Phi_s(t)$ depends on the power allocation
decisions in \emph{both} phases which makes this problem different
from greedy strategies (e.g., \cite{yeh}, \cite{neely-energy}).
In order to determine the optimal strategy in slot $t$, the source $s$ computes
the minimum cost of (\ref{eq:p1ssp3}) for all transmission modes
described earlier and chooses one with the least cost. 

Note that this problem is unconstrained since the long term time average
reliability and power constraints do not appear explicitly as in the
original problem. These are implicitly captured by the virtual queue
values. Further, its solution uses the value of the \emph{current} channel state $\mathcal{T}(t)$ 
and does not require knowledge of the statistics that govern the evolution of the channel state process.
%knowledge of the system parameters (like channel statistics,
%mobility patterns, etc.). 
Thus, the control strategy involves
implementing the solution to the sequence of such unconstrained
problems every slot and updating the queue values according to
(\ref{eq:p1u1}), (\ref{eq:p1x1}).
%Contrast this with static resource allocation policies...
Assuming i.i.d. $\mathcal{T}(t)$ states,
the following theorem characterizes the performance of this dynamic
control algorithm A similar statement
can be made for more general Markov modulated $\mathcal{T}(t)$ using
the techniques of \cite{neely-NOW}. For simplicity, here we consider the i.i.d. case.

\emph{Theorem 1}: (Algorithm Performance) Suppose all queues are
initialized to $0$. Then, implementing the dynamic algorithm
(\ref{eq:p1ssp3}) every slot stabilizes all queues, thereby
satisfying the minimum reliability and time-average power
constraints, and guarantees the following performance bounds (for
some $\epsilon > 0$ that depends on the
slackness of the feasibility constraints):
\begin{align*}
&\lim_{t \rightarrow \infty} \frac{1}{t}
\sum_{\tau=0}^{t-1}\expect{Z_s(\tau)} \leq  \frac{B + V(\alpha_s +
\sum_{i\in \mathcal{\widehat{R}}} \beta_iP_i^{max})}{\epsilon} \\
&\lim_{t \rightarrow \infty} \frac{1}{t}
\sum_{\tau=0}^{t-1}\sum_{i\in \mathcal{\widehat{R}}}\expect{X_i(\tau)}
\leq \frac{B + V(\alpha_s + \sum_{i\in \mathcal{\widehat{R}}}
\beta_iP_i^{max})}{\epsilon}
\end{align*}
Further, the time average utility achieved for any $V \geq 0$
satisfies:
\begin{align*}
\lim_{t \rightarrow \infty} \frac{1}{t} \sum_{\tau=0}^{t-1} \expect{\alpha_s\Phi_s(\tau) - \sum_{i\in \mathcal{\widehat{R}}} \beta_i P_i(\tau)} 
\geq  \zeta  ^* - \frac{B}{V}
%\alpha_s r^*_s- \sum_{i\in \mathcal{\widehat{R}}} \beta_ie^*_i -\frac{B}{V}
\end{align*}
where 
\begin{align*}
& \zeta^* \defequiv \alpha_s r^*_s- \sum_{i\in \mathcal{\widehat{R}}} \beta_ie^*_i \\
& B \defequiv \frac{1 + \lambda_s^2 \rho_s^2 + \sum_{i\in \mathcal{\widehat{R}}} (P_i^{avg})^2 + (P_i^{max})^2}{2}
\end{align*}

%\emph{Proof}: See \cite{urgaonkar-coop}.  \hspace{5.5cm} $\Box$
\emph{Proof}: Appendix A.  \hspace{5.2cm} $\Box$

Thus, one can get within $O(1/V)$ of the optimal values by increasing $V$ at the cost of an $O(V)$ increase 
in the virtual queue backlogs. The size of these queues affects the time required for the time average values to converge to the
desired performance.

%\section{$2$-Stage Resource Allocation Problem}
%\label{section:2stage}

%The optimal solution to (\ref{eq:p1ssp3}) is then given by the minimum cost path in..

%For example, if $s$ uses power
%$P_s(t)$ to transmit directly to $d$, then the packet transmission
%is assumed to be successful if $\mathcal{I}_{sd}(t) = {W} \log\Big(1
%+ \frac{P_s(t)}{W}|h_{sd}(t)|^2\Big) \geq R$.

%Also, we assume $X_s(t) + V, X_i > 0$, else can trivially allocate
%$P_{max}$....Also, if $Z_s = 0$, the zero power allocation trivial
%solution.}

%\footnote{This
%definition is also quite useful in practise as pointed out in
%\cite{caire1}.}

%Note that while the effective $SNR$ increases in the orthogonal schemes, there
%is a loss due to reduced multiplexing gain.

%\underline{\emph{Time-slot Structure}} \emph{illustrate this in Fig.
%\ref{fig:one}. Assuming that $W$ is the total bandwidth available
%for transmission, Fig. \ref{fig:one} shows the slot structure for
%the case of direct transmission to the destination. Figs.
%\ref{fig:2}, \ref{fig:3} show two possible slot structures when an
%orthogonal cooperative protocol is being used. Here, the available
%\textit{degrees of freedom} (e.g. frequency and time) are divided
%among the cooperating nodes so that each one transmits on an
%orthogonal channel. Fig. \ref{fig:4} shows the case in which the
%cooperating nodes use an appropriate distributed space-time code
%(DSTC) so that they can transmit simultaneously on the same channel.}

In the following sections, we investigate the basic $2$-stage resource 
allocation problem (\ref{eq:p1ssp3}) in detail and present solutions for 
two widely studied classes of cooperative protocols proposed in the literature:
Decode-and-Forward (DF) and Amplify-and-Forward (AF)
\cite{Laneman1, Laneman2}. These protocols differ in the way
the transmitted signal from the first phase is processed by the
cooperating relays. In DF, a relay fully decodes the signal. If the
packet is received correctly, it is re-encoded and transmitted in
the second phase. In AF, a relay simply retransmits a scaled version
of the received analog signal. We refer to
\cite{Laneman1, Laneman2} for further details on the working of
these protocols as well as derivation of expressions for the mutual
information achieved by them. Let $m = |\mathcal{R}|$. In the following, we assume a Gaussian
channel model with a total bandwidth $W$ and unit noise power per
dimension. We use the information theoretic
definition of a transmission failure (an outage event) 
as discussed in \cite{tse2}, \cite{caire1}. Here, an outage occurs when the total
instantaneous mutual information is smaller than the rate $R$ at
which data is being transmitted.  

%In what follows, we assume that a packet can be reliably decoded by
%a receiver if the total mutual information exceeds the rate $R$ of
%data transmission. For example, if the source uses power $P_s(t)$,
%then a relay node $R_i$ is successful in decoding if
%$\mathcal{I}_{si}(t) = \frac{W}{m} \log\Big(1 +
%\frac{mP_s(t)}{N_0W}|h_{si}(t)|^2\Big) \geq R$.

%\begin{align}
%\mathcal{I}_{tot} = \frac{W}{m} \log\Big(1 +
%\frac{mP_s}{W}|h_{sd}|^2 + \sum_{i\in \mathcal{U}}
%\frac{mP_i}{W}|h_{id}|^2\Big) \label{eq:mi_dfortho_regen}
%\end{align}

We first consider the case when the channel gains are known at the 
source (Sec. \ref{section:2stage}). In this scenario, (\ref{eq:p1ssp3}) becomes a $2$-stage \emph{deterministic
shortest path problem} because the outcome $\Phi_s(t)$ due to any
control decision and its power allocation can be computed
beforehand. Specifically, $\Phi_s(t) = 1$ when the resulting total
mutual information exceeds $R$ and $\Phi_s(t) = 0$ otherwise. Further, this outcome is a function
of control actions taken over two stages when cooperative transmission is used. This resulting problem is
combinatorial and non-convex and does not admit closed-form solutions in general. 
However, for these protocols, we can reduce it to a set of simpler convex programs for which we can derive
quasi-closed form solutions. Then in Sec. \ref{section:stats}, we consider the case when only the statistics of the channel gains 
are known. In this case, the outcome $\Phi_s(t)$ is random function of the control actions (taken over the two
stages in case of cooperative transmission) and (\ref{eq:p1ssp3}) becomes 
a $2$-stage \emph{stochastic dynamic program}. While standard dynamic programming
techniques can be used to compute the optimal solution, they are typically computationally intensive. 
Therefore, for this case, we present a Monte Carlo simulation based technique to efficiently solve the
resulting dynamic program.

\section{$2$-Stage Resource Allocation Problem with Known Channels, Unknown Statistics}
\label{section:2stage}

Recall that in order to determine the optimal control action in any slot $t$, we must choose between the four
modes of operation as discussed in Sec. \ref{section:basic}: $(1)$  direct transmission, $(2)$
multi-hop relay, $(3)$ cooperative, and $(4)$ idle.  
Let $c_i(t)$ and $I_i(t)$ denote the optimal cost of the metric (\ref{eq:p1ssp3}), and the
corresponding action that achieves that metric, assuming
that mode $i \in \{1, 2, 3, 4\}$ is chosen in slot $t$.  Every slot, the algorithm
computes $c_i(t)$ and $I_i(t)$ for each mode and then implements 
the mode $i$ and the resulting action $I_i(t)$ that
minimizes cost. Note that the cost $c_4(t)$ for the idle
mode is trivially $0$. 
%\underline{\emph{Minimum Cost for Direct Transmission}} 
The minimum cost for direct transmission can be computed as follows.
When the source transmits directly, we have $P_i(t) = 0 \; \forall i \in
\mathcal{R}$. The minimum cost $c_1(t)$ associated with a \emph{successful}
direct transmission ($\Phi_s(t) = 1$) can be obtained by solving the
following convex problem
\footnote{Note that the term $-Z_s(t) - V \alpha_s$ in the objective
is a constant in any given slot and does not affect the solution. 
However, we keep it to compare the net cost between all modes of operation.}: 
\begin{align}
\textrm{Minimize:}\qquad & \Big(X_s(t)+ V \beta_s\Big){P_s(t)} -Z_s(t) - V \alpha_s   \nonumber \\
\textrm{Subject to:} \qquad &{W} \log\Big(1 +
\frac{P_s(t)}{W}|h_{sd}(t)|^2\Big) \geq R \nonumber \\
& 0 \leq P_s(t) \leq P_s^{max}
 \label{eq:p1ssp4}
\end{align}
where the constraint ${W} \log\Big(1 +
\frac{P_s(t)}{W}|h_{sd}(t)|^2\Big) \geq R$ represents the fact that
to get $\Phi_s(t) = 1$, the mutual information must exceed $R$. It
is easy to see that if there is a feasible solution to the above,
then for minimum cost, this constraint must be met with equality.
Using this, the minimum cost corresponding to the direct transmission mode is given by:
$\Big(X_s(t)+ V \beta_s\Big){P_s^{dir}(t)} -Z_s(t) - V\alpha_s$ if $P_s^{dir}(t) = \frac{W}{|h_{sd}(t)|^2}(2^{R/W} - 1) \leq
P_s^{max}$. Otherwise, direct transmission is infeasible and so we set
$c_1(t) = +\infty$.  In this case, direct transmission will not be considered
as the idle mode cost $c_4(t) = 0$ is strictly better, but we must also compare
with the costs $c_2(t)$ and $c_3(t)$.

%\underline{\emph{Minimum Cost for Multi-Hop Transmission}} 

To compute the minimum cost $c_2(t)$ associated with multi-hop transmission, note that
in this case, the slot is divided into two parts (Fig. \ref{fig:one}(b)) and
$P_i(t) > 0$ for at most one $i \in \mathcal{R}$. This strategy is a
special case of the Regenerative DF protocol (to be discussed next)
that uses only $1$ relay and in which the destination does not use
signals received from the first stage for decoding. Therefore, the
optimal cost for this can be calculated using the procedure for the
Regenerative DF case by imposing the single relay constraint and setting $h_{sd}(t) = 0$.
%The cost of modes $1$ and $2$ are also easy to compute. 

Below we present the computation of the minimum cost $c_3(t)$ for the cooperative transmission
mode under several protocols. In what follows, we drop the time subscript $(t)$ for notational convenience.

%\emph{Note: need to know only current channel gains, not statistics.
%Also, channel gains, not phase info... Note that the underlying graph for this SSP is acyclic, so that the
%above problem is essentially a dynamic program.
%Some points:
%\begin{enumerate}
%\item Assumptions on CSI
%\item Protocols considered: AF vs DF
%\item It may be possible to use the ``soft information'' from past
%transmission attempts to be used later. We consider both these
%cases.
%\end{enumerate}
%}

%\underline{{\emph{Example 1: Regenerative DF, Orthogonal Channels}}}
%\subsection{Example 1: Regenerative DF, Orthogonal Channels}
\subsection{Regenerative DF, Orthogonal Channels}
\label{section:df_regen}

Here, the source and relays are each assigned an orthogonal channel
of equal size. An example slot structure is shown in
Fig. \ref{fig:one}(c) in which the entire slot is divided into $m+1$
equal mini-slots.
 In the first phase of the protocol, $s$ transmits
the packet in its slot using power $P_s$. In the second phase, a
subset $\mathcal{U} \subset \mathcal{R}$ of relays that were
successful in reliably decoding the packet, re-encode it using the
\textit{same} code book and transmit to the destination on their
channels with power $P_i$ (where $i \in \mathcal{U}$). 
Given such a set $\mathcal{U}$, the total mutual information 
%$\mathcal{I}_{tot}^{RDF}$
under this protocol
is given by \cite{Laneman1}:
%$\frac{W}{m} \log\Big(1 + \frac{mP_s}{W}|h_{sd}|^2 + \sum_{i\in \mathcal{U}}\frac{mP_i}{W}|h_{id}|^2\Big)$. 
\begin{align*}
%\mathcal{I}_{tot} = 
\frac{W}{m} \log\Big(1 + \frac{mP_s}{W}|h_{sd}|^2 + \sum_{i\in \mathcal{U}} \frac{mP_i}{W}|h_{id}|^2\Big)
\end{align*}
%We can now express (\ref{eq:p1ssp3}) for this protocol as follows.
This is derived by assuming that the receiver uses Maximal Ratio Combining
to process the signals. As seen in the expression for the mutual information, 
such an orthogonal structure increases the SNR,
but utilizes only a fraction of the available degrees of freedom
leading to reduced multiplexing gain.

Define binary variables $x_i$ to be $1$ if relay $i$ can reliably
decode the packet after the first stage and $0$ else. Then, for this
protocol, (\ref{eq:p1ssp3}) is equivalent to the following
optimization problem:
\begin{align}
\textrm{Minimize:}  & (X_s + V \beta_s) P_s + \sum_{i \in \mathcal{R}} (X_i+V \beta_i)P_i - Z_s - V \alpha_s\nonumber \\
\textrm{Subject to:}  & \frac{W}{m} \log\Big(1 + \frac{mP_s}{W}|h_{sd}|^2 + \sum_{i \in \mathcal{R}}x_i\frac{mP_i}{W}|h_{id}|^2\Big) \geq R \nonumber \\
&  \frac{W}{m} \log\Big(1 + \frac{mP_s}{W}|h_{si}|^2\Big) \geq x_iR \nonumber \\
&  0 \leq P_s \leq P_s^{max} \nonumber \\
&  0 \leq  P_i \leq P_i^{max}, x_i\in\{0,1\} \; \forall i \in \mathcal{R} 
\label{eq:dfortho1}
\end{align}

The variables  $x_i$ capture the requirement that a relay can
cooperatively transmit in the second stage only if it was successful
in reliably decoding the packet using the first stage transmission.
A similar setup is considered in \cite{Maric1} but it treats the
limiting case when $W$ goes to infinity.
Because of the integer constraints on $x_i$, (\ref{eq:dfortho1}) is
non-convex. However, we can exploit the structure of this protocol
to reduce the above to a set of $m+1$ subproblems as follows. We
first order the relays in decreasing order of their $|h_{si}|^2$
values. Define $\mathcal{U}_k$ as the set that contains the first
$k$ (where $0 \leq k \leq m$) relays from this ordering. Let
$P_s^{\mathcal{U}_k}$ denote the minimum source power required to
ensure that all relays in $\mathcal{U}_k$ can reliably decode the
packet after the first stage. We note that for all values of $P_s$
in the range $(P_s^{\mathcal{U}_k}, P_s^{\mathcal{U}_{k+1}})$, the
relay set that can reliably decode remains the same, i.e.,
$\mathcal{U}_k$. Thus, we need to consider only $m+1$ subproblems,
one for each $\mathcal{U}_k$. The subproblem for any set
$\mathcal{U}_k$ is given by:
%$\mathcal{U}_k$ can be expressed as an LP:
\begin{align}
\textrm{Minimize:} \; & (X_s + V \beta_s) P_s + \sum_{i\in \mathcal{U}_k} (X_i+V \beta_i)P_i - Z_s- V \alpha_s\nonumber \\
\textrm{Subject to:} \; & \frac{W}{m} \log\Big(1 + \frac{mP_s}{W}|h_{sd}|^2 + \sum_{i\in \mathcal{U}_k} \frac{mP_i}{W}|h_{id}|^2\Big) \geq R \nonumber \\
& P_s^{\mathcal{U}_k} \leq P_s \leq P_s^{max} \nonumber \\
& 0 \leq P_i \leq P_i^{max} \qquad \forall i \in \mathcal{U}_k
\label{eq:dfortho2}
\end{align}
This can easily be expressed as the following LP:
\begin{align}
\textrm{Minimize:} \; & (X_s + V \beta_s) P_s + \sum_{i\in \mathcal{U}_k}  (X_i+V \beta_i)P_i - Z_s- V \alpha_s\nonumber \\
\textrm{Subject to:} \; & P_s|h_{sd}|^2 + \sum_{i\in \mathcal{U}_k}P_i|h_{id}|^2 \geq \theta \nonumber \\
& P_s^{\mathcal{U}_k} \leq P_s \leq P_s^{max} \nonumber \\
& 0 \leq P_i \leq P_i^{max} \qquad \forall i \in \mathcal{U}_k
\label{eq:dfortho3}
\end{align}
where $\theta = \frac{W}{m} (2^{Rm/W} - 1)$. The solution to the LP 
above has a greedy structure where we start by allocating increasing
power to the nodes (including $s$) in decreasing order of the value
of $\frac{|h_{id}|^2}{(X_i+V\beta_i)}$ (where $i \in \mathcal{U}_k \cup \{s\}$)
 till any constraint is met. 

Therefore, for this protocol, the optimal solution to finding the
cost $c_3(t)$ associated with the cooperative transmission mode in
(\ref{eq:p1ssp3}) can be computed by solving (\ref{eq:dfortho3}) for
each $\mathcal{U}_k$ and picking the one with the least cost. 
It is interesting to note that if we impose a constraint on the sum total
power of the relays instead of individual node constraints, then due to the greedy
nature of the solution to (\ref{eq:dfortho3}), it is optimal to
select at most $1$ relay for cooperation. Specifically, this relay is the one that has the
highest value of $\frac{|h_{id}|^2}{(X_i+V\beta_i)}$.

%\emph{If also optimizing over slot structure, can show using concavity of log function that this is also optimal 
%over all slot structures.}

\subsection{Non-Regenerative DF, Orthogonal Channels}
\label{section:df_nonregen}

This protocol is similar to Regenerative DF protocol discussed in Sec. \ref{section:df_regen}. 
The only difference is that here, in the second stage,
the subset $\mathcal{U} \subset \mathcal{R}$ relays that were
successful in reliably decoding the packet re-encode it using
\textit{independent} code books. In this case, the total mutual
information is given by \cite{Laneman2}:
%$\frac{W}{m}\log\Big(1 + \frac{mP_s}{W}|h_{sd}|^2\Big) + \sum_{i\in \mathcal{R}} \frac{W}{m}\log\Big(1 + x_i\frac{mP_i}{W}|h_{id}|^2\Big)$.
\begin{align*}
\frac{W}{m}\log\Big(1 + \frac{mP_s}{W}|h_{sd}|^2\Big) + \sum_{i\in
\mathcal{R}} \frac{W}{m}\log\Big(1 +
x_i\frac{mP_i}{W}|h_{id}|^2\Big)
\end{align*}
Using the same definition of binary variables $x_i$ as in Sec.\ref{section:df_regen} , we can
express (\ref{eq:p1ssp3}) for this protocol as an optimization
problem that resembles (\ref{eq:dfortho1}). Similar to the
Regenerative DF case, we can then reduce this to a set of $m+1$
subproblems, one for each $\mathcal{U}_k$.
The subproblem
for set $\mathcal{U}_k$ is given by:
\begin{align}
&\textrm{Minimize:}\; \; (X_s + V \beta_s) P_s + \sum_{i\in \mathcal{U}_k} (X_i+V \beta_i)P_i - Z_s- V \alpha_s \nonumber \\
&\textrm{Subject to:} \nonumber \\
 &\log\Big(1 +
\frac{mP_s}{W}|h_{sd}|^2\Big) + \sum_{i\in \mathcal{U}_k} \log\Big(1
+ \frac{mP_i}{W}|h_{id}|^2\Big)
 \geq \frac{mR}{W}\nonumber \\
& P_s^{\mathcal{U}_k} \leq P_s \leq P^{max} \nonumber \\
& 0 \leq P_i \leq P^{max} \qquad \forall i \in \mathcal{U}_k
\label{eq:nonregen_dfortho2}
\end{align}
The above problem is convex and we can use the KKT
conditions to get the optimal solution (see Appendix B for details). 
Define $[x]_0^{P^{max}}
\defequiv \min[\max(x, 0), P^{max}]$. Then the solution to the subproblem for
set $\mathcal{U}_k$ is given by:
\begin{align}
&P_s^*(\mathcal{U}_k) = \Big[\frac{\nu^*}{X_s+V\beta_s} - \frac{W}{m|h_{sd}|^2}\Big]_{P_s^{\mathcal{U}_k}}^{P_s^{max}} \nonumber \\ 
& P_i^*(\mathcal{U}_k) = \Big[\frac{\nu^*}{X_i+V\beta_i} - \frac{W}{m|h_{id}|^2}\Big]_0^{P_i^{max}} \forall i \in \mathcal{U}_k
\label{eq:nonregen_dfortho_sol}
\end{align}
where $\nu^* \geq 0$ is chosen so that the total mutual information
constraint is met with equality. 
Therefore, the optimal solution for the cost $c_3(t)$ in (\ref{eq:p1ssp3}) for this
protocol can be computed by solving (\ref{eq:nonregen_dfortho_sol})
for each $\mathcal{U}_k$ and picking one with the least cost.
We note that the solution above has a water-filling type structure that is typical of related resource allocation problems in
static settings.

%In Appendix A, we use the KKT conditions to solve the above convex
%optimization problem. Define $[x]_0^{P_{max}}
%\defequiv \min[\max(x, 0), P_{max}]$. The solution is given by:

%\emph{Some points:
%1. If constraint on sum total power instead of individual, it is not
%if also optimizing over slot structure, then at most $1$ relay
%optimal over all slot structures.}

\subsection{AF, Orthogonal Channels} 
\label{section:af_regen}

In this protocol, the source and relays are again assigned an orthogonal
channel of equal size. An example slot structure is shown in
Fig. \ref{fig:one}(c). However, instead of trying to decode the packet, the
relays amplify and forward the received signal from the first stage.
The total mutual information under this protocol is given by
\cite{Maric2} \cite{Adve1}:
%$\frac{W}{m} \log\Bigg(1 +
%\frac{mP_s}{W}\Big(|h_{sd}|^2 + \sum_{i \in \mathcal{R}} \psi_i
%\Big)\Bigg)$
\begin{align*}
%\mathcal{I}_{tot} = 
\frac{W}{m} \log\Bigg(1 + \frac{mP_s}{W}\Big(|h_{sd}|^2 + \sum_{i \in \mathcal{R}} \psi_i \Big)\Bigg)
\end{align*}
%\begin{align}
%\mathcal{I}_{tot} = \frac{W}{m} \log\Bigg(1 +
%\frac{mP_s}{W}\Big[|h_{sd}|^2 + \sum_{i \in \mathcal{R}_{s}} \psi_i
%\Big]\Bigg) \label{eq:mi_afortho}
%end{align}
where $\psi_i \defequiv \frac{P_i|h_{si}|^2|h_{id}|^2}{P_s|h_{si}|^2 + P_i|h_{id}|^2 + W/m}$. 
Using this, we can express (\ref{eq:p1ssp3}) for this model as follows.
\begin{align}
\textrm{Minimize:} \; & (X_s + V \beta_s) P_s + \sum_{i \in \mathcal{R}} (X_i+V \beta_i)P_i  - Z_s - V \alpha_s\nonumber \\
\textrm{Subject to:} \;  & \frac{W}{m} \log\Bigg(1 + \frac{mP_s}{W}\Big(|h_{sd}|^2 + \sum_{i \in \mathcal{R}} \psi_i \Big)\Bigg)\geq R \nonumber \\
 & 0 \leq P_s \leq P_s^{max} \nonumber \\
 & 0 \leq P_i \leq P_i^{max} \; \forall i \in \mathcal{R}
\label{eq:afortho1}
\end{align}
This problem is non-convex. However, if we fix the source power
$P_s$, then it becomes convex in the other variables. This reduction
has been used in \cite{Adve1} as well, although it considers a
static scenario with the objective of minimizing instantaneous
outage probability. After fixing $P_s$, we can compute the optimal
relay powers for this value of $P_s$ by solving the following:
\begin{align}
\textrm{Minimize:} \qquad & \sum_{i \in \mathcal{R}} (X_i+V \beta_i)P_i - Z_s - V \alpha_s\nonumber \\
\textrm{Subject to:} \qquad & P_s|h_{sd}|^2 + \sum_{i \in
\mathcal{R}} P_s\psi_i \geq \theta \nonumber \\
\qquad &0 \leq P_i \leq P_i^{max} \qquad \forall i \in \mathcal{R}
\label{eq:afortho2}
\end{align}
where $\theta = \frac{W}{m} (2^{Rm/W} - 1)$. The first constraint can be simplified as:
%\begin{align*}
%P_s|h_{sd}|^2 + \sum_{i \in \mathcal{R}} P_s\psi_i =  P_s(|h_{sd}|^2 + \sum_{i \in \mathcal{R}} |h_{si}|^2) 
%\\ - \sum_{i \in \mathcal{R}} \frac{P_s^2|h_{si}|^4 + P_s|h_{si}|^2W/m}{P_s|h_{si}|^2 + P_i|h_{id}|^2 +W/m}
%\end{align*}

$P_s|h_{sd}|^2 + \sum_{i \in \mathcal{R}} P_s\psi_i =  P_s(|h_{sd}|^2 + \sum_{i \in \mathcal{R}} |h_{si}|^2) 
 - \sum_{i \in \mathcal{R}} \frac{P_s^2|h_{si}|^4 + P_s|h_{si}|^2W/m}{P_s|h_{si}|^2 + P_i|h_{id}|^2 +W/m}$

Since we have fixed $P_s$, we can express (\ref{eq:afortho2}) as:
\begin{align}
\textrm{Minimize:} \qquad & \sum_{i \in \mathcal{R}} (X_i+V \beta_i)P_i - Z_s - V \alpha_s\nonumber \\
\textrm{Subject to:} \qquad & \sum_{i\in
\mathcal{R}}\frac{P_s^2|h_{si}|^4 +
P_s|h_{si}|^2W/m}{P_s|h_{si}|^2 + P_i|h_{id}|^2 +W/m} \leq \theta' \nonumber \\
\qquad &0 \leq P_i \leq P_i^{max} \qquad \forall i \in
\mathcal{R}\label{eq:afortho3}
\end{align}
where $\theta' = P_s(|h_{sd}|^2 + \sum_{i \in \mathcal{R}_{s}}
|h_{si}|^2) - \theta$.
%In Appendix B, we use the KKT conditions to
%solve the above convex optimization problem. The solution is given
%y:
Using the KKT conditions, the solution the above convex
optimization problem is given by (see Appendix C for details):
%\begin{align*}
$P_i^* = \Big[\sqrt{\frac{\nu^*(P_s^2|h_{si}|^4 + P_s|h_{si}|^2W/m)}{(X_i+V\beta_i)|h_{id}|^2}} 
- \frac{P_s|h_{si}|^2+ W/m}{|h_{id}|^2}\Big]_0^{P_i^{max}}$
%\end{align*}
where $\nu^* \geq 0$ is chosen so that the second constraint is met with equality. We note that this
solution has a water-filling type structure as well. Therefore, to compute 
the optimal solution to (\ref{eq:p1ssp3}) for
this protocol, we would have to solve the above for each value of
$P_s \in [0, P_s^{max}]$. In practice, this computation can be
simplified by considering only a discrete set of values for $P_s$.
Because we have derived a simple closed form expression
for each $P_s$, it is easy to compare these values over, say,
a discrete list of $100$ options in $[0, P_s^{max}]$ to pick the best
one, which enables a very accurate approximation to optimality in
real time.

\subsection{DF with DSTC}
\label{section:df_dstc}

In this protocol, all the cooperating relays in the second stage use an appropriate
distributed space-time code (DSTC) \cite{Laneman2} so that they can
transmit simultaneously on the same channel. The slot structure
under this scheme is shown in Fig.\ref{fig:one}(d). 
Suppose in the first phase of the protocol, $s$ transmits
the packet in the first half of the slot  using power $P_s$. In the second phase, a
subset $\mathcal{U} \subset \mathcal{R}$ of relays that were
successful in reliably decoding the packet, re-encode it using a DSTC 
and transmit to the destination with power $P_i$ (where $i \in \mathcal{U}$) in the second half of the slot. 
Given such a set $\mathcal{U}$, the total mutual information 
%$\mathcal{I}_{tot}$
under this protocol is given by \cite{Laneman1}: 
%$\frac{W}{2} \log\Big(1 + \frac{2P_s}{W}|h_{sd}|^2 + \sum_{i\in \mathcal{U}} \frac{2P_i}{W}|h_{id}|^2\Big)$. 
\begin{align*}
%\mathcal{I}_{tot} = 
\frac{W}{2} \log\Big(1 + \frac{2P_s}{W}|h_{sd}|^2 + \sum_{i\in \mathcal{U}} \frac{2P_i}{W}|h_{id}|^2\Big)
\end{align*}
 The factor of $2$ appears because only half of the slot is being
used for transmission. As seen in the expression above, unlike the earlier examples, this protocol does not suffer from reduced
multiplexing gains due to orthogonal channels.

We can now express (\ref{eq:p1ssp3}) for this protocol as follows.
Define binary variables $x_i$ to be $1$ if relay $i$ can reliably
decode the packet after the first stage and $0$ else. Then, for this
protocol, (\ref{eq:p1ssp3}) is equivalent to the following
optimization problem:
\begin{align}
\textrm{Minimize:} \; & (X_s + V \beta_s) P_s + \sum_{i \in \mathcal{R}} (X_i+V \beta_i)P_i - Z_s - V \alpha_s\nonumber \\
\textrm{Subject to:} \; & \frac{W}{2} \log\Big(1 + \frac{2P_s}{W}|h_{sd}|^2 + \sum_{i \in \mathcal{R}}x_i\frac{2P_i}{W}|h_{id}|^2\Big) \geq R \nonumber \\
& \frac{W}{2} \log\Big(1 + \frac{2P_s}{W}|h_{si}|^2\Big) \geq x_iR \nonumber \\
&  0 \leq P_s \leq P_s^{max} \nonumber \\
&  0 \leq  P_i \leq P_i^{max}, x_i\in\{0,1\} \; \forall i \in \mathcal{R} 
\label{eq:dfdstc1}
\end{align}
By comparing the above with (\ref{eq:dfortho1}), it can be seen that the 
computation of minimum cost under this protocol follows the same procedure
as described in Sec. \ref{section:df_regen} of solving $m+1$ subproblems, each an LP, by ordering the relays greedily
and hence we do not repeat it.

%To solve (\ref{eq:dfdstc1}), we consider a set of $m$ subproblems as
%follows. We first order the relays in decreasing order of their
%$|h_{si}(t)|^2$ values. Define set $s_k$ as the set that contains
%the first $k$ relays from this ordering. Let $P_s^{s_k}(t)$ denote
%the minimum source power required to ensure that all relays in the
%set $s_k$ can reliably decode the packet after the first stage.
%Then, the subproblem for this set $s_k$ is given by:

%The solution to the
%above has a greedy structure where we start by allocating increasing
%power to the node in decreasing order of the value of
%$\frac{|h_{id}|^2}{X_i}$ till any constraint is met.

%Some points: 
%1. If constraint on sum total power instead of
%individual, optimal to select at most $1$ relay for cooperation.
%Could get a closed form for which relay to choose if this.

\subsection{AF with DSTC}
\label{section:af_dstc}

Here, all cooperating relays use amplify and forward along with DSTC.
The total mutual information under this protocol is given by:
\begin{align*}
%\mathcal{I}_{tot} = 
\frac{W}{2} \log\Bigg(1 + \frac{2P_s}{W}\Big(|h_{sd}|^2 + \sum_{i \in \mathcal{R}} \psi_i \Big)\Bigg)
\end{align*}
where $\psi_i = \frac{P_i|h_{si}|^2|h_{id}|^2}{P_s|h_{si}|^2 + P_i|h_{id}|^2 + W/2}$. 
Using this, we can express (\ref{eq:p1ssp3}) for this model as follows.
\begin{align}
\textrm{Minimize:} \; & (X_s + V \beta_s) P_s + \sum_{i \in \mathcal{R}} (X_i+V \beta_i)P_i  - Z_s - V \alpha_s\nonumber \\
\textrm{Subject to:} \;  & \frac{W}{2	} \log\Bigg(1 + \frac{mP_s}{W}\Big(|h_{sd}|^2 + \sum_{i \in \mathcal{R}} \psi_i \Big)\Bigg)\geq R \nonumber \\
 & 0 \leq P_s \leq P_s^{max} \nonumber \\
 & 0 \leq P_i \leq P_i^{max} \; \forall i \in \mathcal{R}
\label{eq:afdstc1}
\end{align}
This is similar to (\ref{eq:afortho1}) and thus, we fix $P_s$ and use a similar reduction to get
a convex optimization problem whose solution can be derived using KKT conditions and is
given by:

$P_i^* = \Big[\sqrt{\frac{\nu^*(P_s^2|h_{si}|^4 + P_s|h_{si}|^2W/2)}{(X_i+V\beta_i)|h_{id}|^2}} 
- \frac{P_s|h_{si}|^2+ W/2}{|h_{id}|^2}\Big]_0^{P_i^{max}}$
%\end{align*}
where $\nu^* \geq 0$ is chosen so that the constraint on the total mutual information at the destination 
is met with equality. 

%To compute the optimal solution to (\ref{eq:p1ssp3}) for
%this protocol, we would have to solve the above for each value of
%$P_s \in [0, P_s^{max}]$. In practice, this computation can be
%simplified by considering only a discrete set of values for $P_s$.

\section{$2$-Stage Resource Allocation Problem with Unknown Channels, Known Statistics}
\label{section:stats}

We next consider the solution to (\ref{eq:p1ssp3}) when the source
does not know the current channel gains and is only aware of their
statistics. In this case, (\ref{eq:p1ssp3}) becomes a
$2$-stage stochastic dynamic program. For brevity, here we focus on its solution for the
cooperative transmission mode.

 Suppose the source uses power $P_s$ in the first stage.
Let $\omega$ denote the outcome of this transmission. This lies in a
space $\Omega$ of possible network states which is assumed to be of a finite
but arbitrarily large size. For
example, in the DF protocol, $\omega$ might represent the set of relay
nodes that received the packet successfully after the first stage as well as the mutual
information accumulated so far at the destination. 
%It can be seen that invthe $2$-stage problem with $m$ relays, there can be $2^m$ differentcoutcomes.
 For AF,  $\omega$ can represent the SNR value at each
relay node and at the destination.
% The space for this will be a subset of $\mathbb{R}^m$ and will have infinite elements. 
%For simplicity, here we consider
%the DF protocol. 

Let $J_1^*(P_s, \omega)$ be the optimal cost-to-go
function for the $2$-stage dynamic program (\ref{eq:p1ssp3}) given that the source uses power $P_s$ in the
first stage and the network state is $\omega$ at the beginning of the second stage. Let $J_0^*$ denote
the optimal cost-to-go function starting from the first stage. Also, let $\mathcal{R}(\omega)$ denote 
the set of relay nodes that can take part in cooperative transmission when the network state in $\omega$.
We define the following probabilities. Let $f(P_s, \omega)$ be the
probability that the outcome of the first stage is $\omega$ when the
source uses power $P_s$. Also, let $g(\overrightarrow{P}_{\mathcal{R}(\omega)},
P_s, \omega)$ be the probability that the receiver gets the packet
successfully when relays in $\mathcal{R}(\omega)$ use a power allocation
$\overrightarrow{P}_{\mathcal{R}(\omega)}$ and the source uses power $P_s$. Note
that these probabilities are obtained by taking expectation over all
channel state realizations. We assume these are obtained from the
knowledge of the channel statistics.

Using these definitions, we can now write the Bellman optimality
equations \cite{bertsekas} for this dynamic program $\forall \omega
\in \Omega$:
\begin{align}
&J_0^* = \min_{P_s} \Big[(X_s + V \beta_s) P_s + \sum_{\omega \in \Omega}f(P_s,\omega)J_1^*(P_s, \omega)\Big] \label{eq:dp1} \\
&J_1^*(P_s, \omega) = \min_{\overrightarrow{P}_{\mathcal{R}(\omega)}} \Big[\sum_{i \in \mathcal{R}(\omega)} (X_i + V \beta_i) P_i \nonumber \\ 
&\qquad \qquad \qquad - (Z_s + V \alpha_s) g(\overrightarrow{P}_{\mathcal{R}(\omega)},P_s, \omega)\Big]  
\label{eq:dp2}
\end{align}

While this can be solved using standard dynamic programming
techniques, it has a computational complexity that grows with the
state space size $\Omega$ and can be prohibitive when this is large.
We therefore present an alternate method based on the idea of
Monte Carlo simulation.

%\underline{\emph{Simulation Based Method}}: 
\subsection{Simulation Based Method} 

Suppose the transmitter
performs the following simulation. Fix a source power $P_s$. Define $J^*_0(P_s)$ as the optimal 
cost-to-go function \emph{given} that the source uses power $P_s$.
Note that this is simply the expression on the right hand side of (\ref{eq:dp1}) with $P_s$ fixed.
Simulate the outcome of a transmission at this power $n$ times independently using
the values of $f(P_s, \omega)$. Let $\omega_j \in \Omega$ denote the
outcome of the $j^{th}$ simulation. For each generated outcome
$\omega_j$, compute the optimal cost-to-go function $J_1^*(P_s,
\omega_j)$ by solving (\ref{eq:dp2}) (this could be done using the
knowledge of $g(\overrightarrow{P}_{\mathcal{R}_(\omega)}, P_s, \omega)$ either
analytically or numerically). Use this to update $J_0^{est}(P_s,
n)$, which is an \emph{estimate} of $J^*_0(P_s)$ for a given $P_s$ after $n$ iterations and is defined as follows:
\begin{align}
J_0^{est}(P_s, n) = (X_s + V \beta_s)P_s + \frac{1}{n} \sum_{j=1}^n
J_1^*(P_s, \omega_j) \label{eq:est1}
\end{align}

We now show that, for a given $P_s$, $J_0^{est}(P_s, n)$ can be
pushed arbitrarily close to the optimal cost-to-go function
$J_0^*(P_s)$ by increasing $n$. Since we have fixed $P_s$, from
(\ref{eq:dp1}), we have:
\begin{align*}
&J_0^*(P_s) = (X_s + V \beta_s) P_s + \sum_{\omega \in \Omega}f(P_s,
\omega)J_1^*(P_s, \omega)
\end{align*}

Define the following indicator random variables for each simulation
$j$ and $\forall \omega \in \Omega$:
\begin{displaymath}
1_{\omega}(P_s, j) = \left\{ \begin{array}{lll} 1 & \textrm{if the outcome of simulation $j$ is $\omega$}\\
0 & \textrm{else}
\end{array} \right.
\end{displaymath}

Note that by definition $\expect{1_{\omega}(P_s, j)} = f(P_s,
\omega)$. Therefore, we can express $J_0^{est}(P_s, n)$ in terms of
these indicator variables as follows:
\begin{align*}
J_0^{est}(P_s, n) = &(X_s + V \beta_s) P_s + \frac{1}{n} \sum_{j=1}^n  \sum_{\omega \in \Omega} 1_{\omega}(P_s,j)J_1^*(P_s, \omega)
%\label{eq:est2}
\end{align*}

We note that $\Big(\sum_{\omega \in \Omega}
1_{\omega}(P_s,j)J_1^*(P_s, \omega)\Big)$ are i.i.d. random
variables with mean $\mu = \sum_{\omega \in \Omega}f(P_s,
\omega)J_1^*(P_s, \omega)$ and variance $\sigma^2 = \sum_{\omega \in
\Omega}f(P_s, \omega)(J_1^*(P_s, \omega))^2 - \mu^2$. Using
Chebyshev's inequality, we get for any $\epsilon > 0$:
\begin{align*}
Pr\Big[|\frac{1}{n} \sum_{j=1}^n \Big(\sum_{\omega \in \Omega}
1_{\omega}(P_s,j)J_1^*(P_s, \omega)\Big) - \mu| \geq \epsilon \Big]
\leq \frac{\sigma^2}{n\epsilon^2}
\end{align*}

This shows that the value of the estimate quickly converges 
to the optimal cost-to-go value. Thus, this method can be used to get a good estimate of
the optimal cost-to-go function for a fixed value of $P_s$ in a reasonable number of steps.

\section{Multi-Source Extensions}
\label{section:extensions}

In this section, we extend the basic model of Sec. \ref{section:basic} to the case when there are multiple sources in the network. Let the set of 
source nodes be given by $\mathcal{S}$. We consider the case when all source nodes have orthogonal channels.\footnote{For the non-orthogonal 
scenario, there will two sources of outages: transmission failure at the physical layer and delay violation due to contention in medium access. 
Hence, MAC scheduling in addition to physical layer resource allocation must be considered. This is not the focus of the current work.}
%In the basic model, with all sources having orthogonal  channels, no outages due to MAC contention.}
In particular, we assume that in each slot, a medium access process $\chi(t)$ determines which source nodes get transmission opportunities. 
For simplicity, we assume that at most one source transmits in a slot. This models situations where there might be a pseudo-random TDMA schedule 
that  determines a unique transmitter node every slot. It also models situations where the source nodes use a contention-resolution mechanism such 
as CSMA. Our model can be extended to scenarios where more than one source node can transmit, potentially over orthogonal 
frequency channels.

Let $s(t) = s(\chi(t)) \in \mathcal{S}$ be the source node that gets a transmission opportunity in slot $t$. Then, 
the optimal resource allocation framework developed in Sec. \ref{section:CNC} can be applied as follows. A virtual 
reliability queue is defined for each source node $s \in \mathcal{S}$ and is updated as in (\ref{eq:p1u1}). Note 
that in slots where a source node $s$ does not get a transmission opportunity, $\Phi_s(t) = 0$. We assume that each 
incoming packet gets one transmission opportunity so that the delay constraint of $1$ slot per packet only measures 
the transmission delay and not the queueing delay that would be incurred due to contention. Similarly, a virtual 
power queue is maintained for each node as in (\ref{eq:p1x1}) including the source nodes and relay nodes. Note 
that in this model, it is possible for a source node to act as a relay for another source node when it is not 
transmitting its own data. We denote the set of relay nodes (that includes such source nodes) in slot $t$ as 
$\mathcal{R}(t)$.

Then the optimal control algorithm operates as follows. Let $\textbf{\emph{Q}}(t)$ denote the collection of all virtual queues in 
timeslot $t$. Every slot, given $\textbf{\emph{Q}}(t)$ and any channel state $\mathcal{T}(t)$, it chooses a control action 
$\mathcal{I}_{s(t)}$ that minimizes the following stochastic metric (for a given control parameter $V \geq 0$): 
\begin{align} 
\textrm{Minimize:} \qquad & (X_{s(t)} +V\beta_{s(t)})\expect{P_{s(t)}|\textbf{\emph{Q}}(t), \mathcal{T}(t)} \nonumber \\
&+ \sum_{i \in \mathcal{R}(t)}(X_i(t)+V\beta_i)\expect{P_i(t)|\textbf{\emph{Q}}(t), \mathcal{T}(t)}\nonumber \\
&- (Z_{s(t)}+V \alpha_{s(t)})\expect{\Phi_{s(t)}|\textbf{\emph{Q}}(t), \mathcal{T}(t)} \nonumber \\
\textrm{Subject to:} \qquad & 0 \leq P_{s(t)} \leq P_{s(t)}^{max} \nonumber \\
 			& 0 \leq P_i(t) \leq P_i^{max} \; \forall i \in \mathcal{R}(t) \nonumber \\
			& \mathcal{I}_{s(t)} \in \mathcal{C}
\label{eq:p1ssp3_ms}
\end{align}
This problem can be solved using the techniques described for the single source case.

%TDMA based round-robin schedule exists and each source is assigned a unique slot for transmission in every frame 
%of this schedule.\footnote{Our analysis can be extended to the scenario where the channels are orthogonal in the 
%frequency domain.} Then, the optimal resource framework developed in Sec. \ref{section:CNC} can be applied as 
%follows. Instead of using the virtual queue values at each slot, the values at the \emph{beginning} of a frame are 
%used for all slots in that frame.

%Note: sources can act as relays for other sources. Set of relays can change over time. Captures scenarios with
%network dynamics with mobility.

%This could be achieved, for
%example, by assigning every source a unique transmission opportunity
%in each frame of a TDMA schedule, round robin over all sources. 

\section{Simulations}
\label{section:sim}

\begin{figure}[t]
\centering
\includegraphics[width=8cm,  angle=0]{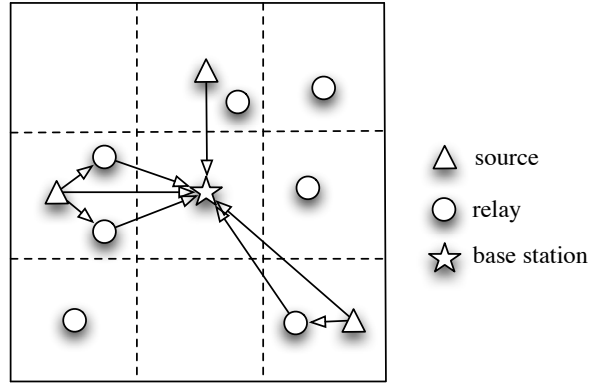}
\caption{A snapshot of the example network used in simulation.}
\label{fig:cell}
\end{figure}

We simulate the dynamic control algorithm (\ref{eq:p1ssp3}) in an
ad-hoc network with $3$ stationary sources and $7$ mobile relays
as shown in Fig. \ref{fig:cell}. Every slot, the sources receive new packets
destined for the base station according to an i.i.d. Bernoulli
process of rate $\lambda$ and each packet has a delay constraint of
$1$ slot. The sources are assumed to have orthogonal channels and
can transmit either directly or cooperatively with a subset of the
relays in their vicinity. We impose a cell-partitioned structure so
that a source can only cooperate with the relays that are in the
same cell in that slot. The relays move from one cell to the other
according to a Markovian random walk. In the simulation, at the end
of every slot, a relay decides to stay in its current cell with
probability $0.8$, else decides to move to an adjacent cell with
probability $0.2$ (where any of the feasible adjacent cells are
equally likely).

We assume a Rayleigh fading model. The amplitude squares of the
instantaneous gains on the links involving a source, the set of
relays in its cell in that slot and the base station are
exponentially distributed random variables with mean $1$. All power
values are normalized with respect to the average noise power. All
nodes have an average power constraint of $1$ unit and a maximum
power constraint of $10$ units.

\begin{figure}
\centering
\includegraphics[width=8cm,  angle=0]{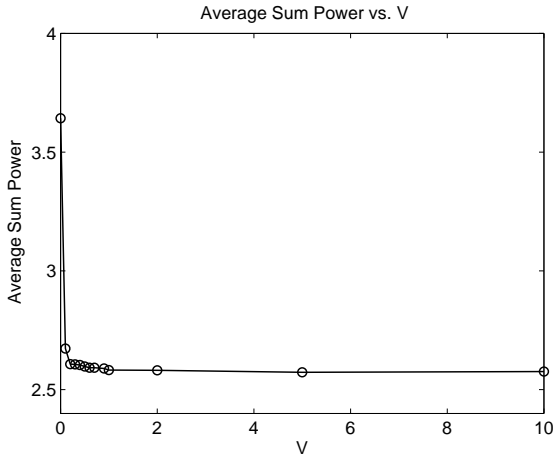}
\caption{Average Sum Power vs. V.}\label{fig:fig3b}
\end{figure}

We consider the Regenerative DF cooperative protocol over orthogonal
channels and implement the optimal resource allocation strategy as
computed in (\ref{eq:dfortho3}) for this network. In the first
experiment, we consider the objective of minimizing the average sum
power expenditure in the network given a minimum reliability
constraint $\rho_s = 0.98$ and input rate $\lambda_s = 0.5$
packets/slot for all sources. For this, we set $\alpha_s =0$ and
$\beta_i = 1$. Fig. \ref{fig:fig3b} shows the average sum power for
different values of the control parameter $V$. It is seen that this
value converges to $2.6$ units for increasing values of $V$, as predicted
by the performance bounds on the time average utility in Theorem
$1$. Fig. \ref{fig:fig3c} shows the resulting average reliability
queue occupancy. It is seen to increase linearly in $V$, again as
predicted by the bound on the time average queue backlog in Theorem
$1$.
We emphasize again
that there are no actual queues in the system, and all successfully delivered
packets have a delay exactly equal to $1$ slot.  The fact that all reliability queues
are stable ensures that we are indeed meeting or exceeding
the $98\%$ reliability constraint.  Indeed, in our simulations we found reliability
to be almost exactly equal to the $98\%$ constraint, as expected in an algorithm
designed to minimize average power subject to this constraint.
We further note that the instantaneous reliability queue value $Z(t)$ represents
the worst case ``excess'' packets that did not meet the reliability constraints over any
interval ending at time $t$,  so that maintaining small $Z(t)$ (with a small $V$)
makes the timescales over which the time average reliability constraints are
satisfied smaller.

In the second experiment, we choose both $\alpha_s =0$ and $\beta_i
= 0$ so that (\ref{eq:obj1}) becomes a feasibility problem. We fix
the average and peak power values to $1$ and $10$ respectively and
implement (\ref{eq:dfortho3}) for different rate-reliability pairs.
In Table \ref{table:notation}, we show whether these are feasible or not under three resource allocation
strategies: direct transmission, always cooperative transmission and dynamic cooperation (that corresponds to
implementing the solution to (\ref{eq:dfortho3}) every slot). It can be seen that dynamic cooperation
significantly increases the feasible rate-reliability region over direct transmission as well as
static cooperation.
For example, it is impossible to achieve
$95\%$ reliability using direct transmission alone, even if the traffic
rate is only $0.2$ packets/slot.  This can be achieved by an algorithm that uses
the cooperation mode (mode $3$) always, but optimizes over the
power allocation decisions of this cooperation mode as specified
in previous sections.  However, always using cooperation fails
if we desire $98\%$ reliability, but using our optimal policy that dynamically
mixes between the different modes, and chooses efficient power
allocation decisions in each mode, can achieve $98\%$ reliability, even
at increased rates up to $0.6$ packets/slot.

\begin{center}
\begin{table*}
{\small
\hfill{}
%\centering
  \begin{tabular}{| c | c | c | c | c | c | c | c |}
   \hline
    (rate, reliability) = $(\lambda_s, \rho_s$) & (0.1, 0.9) & (0.2, 0.9) & (0.2, 0.95) & (0.5, 0.95) & (0.5, 0.98) & (0.6, 0.98) &(0.7, 0.99) \\ \hline
    direct transmission & \checkmark & \checkmark & \textbf{x} & \textbf{x} & \textbf{x} & \textbf{x} & \textbf{x}\\ \hline
    always cooperate & \checkmark & \checkmark & \checkmark & \checkmark & \textbf{x} & \textbf{x} & \textbf{x}\\ \hline
    optimal strategy & \checkmark & \checkmark & \checkmark & \checkmark & \checkmark & \checkmark & \textbf{x}\\ \hline
  \end{tabular}}
\hfill{}
\caption{Table showing the feasibility of different rate-reliability pairs.} 
\label{table:notation}
\end{table*}
\end{center}

%\subsection{$D\leq K$, no past soft information}

%\subsection{$D\leq K$, could use past soft information}

%\subsection{Mobile relays}

\section{Conclusions}
\label{section:conclu}

In this paper, we considered the problem of optimal resource
allocation for delay-limited cooperative communication in a
mobile ad-hoc network.  Using the technique of Lyapunov optimization, 
we developed dynamic cooperation strategies that make
optimal use of network resources to achieve a target outage
probability (reliability) for each user subject to average power
constraints. Our framework is general enough to be applicable
to a large class of cooperative protocols. In particular, in this paper,
we derived  quasi-closed form solutions for several variants of the
Decode-and-Forward and Amplify-and-Forward strategies.

%There are several possible extensions to this work 

%optimizing slot structure

%allowing relays to listen to each other

%alternatives to  ``transmitter'' driven model 

%(where the transmitter does the required computation and determines the control actions), 
%alternative models can be considered (say if they make
%more sense from the point of implementation). For example, one
%option could be, we only optimize relay powers. Here, source will
%just send with a fixed power. Then, depending on the outcome, the
%destination may determine what the relays should do.

%\appendices
\section*{Appendix A: Proof of Theorem $1$}
Here, we prove Theorem $1$ by comparing the Lyapunov drift of the
dynamic control algorithm (\ref{eq:p1ssp3}) with that of an optimal
stationary, randomized policy. Let $r^*_s$ and $e^*_i \; \forall i
\in \mathcal{\widehat{R}}$ denote the optimal value of the objective in
(\ref{eq:obj1}). Then we have the following fact\footnote{This
can be shown using the techniques developed in \cite{neely-energy}.}:

\begin{figure}
\centering
\includegraphics[width=8cm,  angle=0]{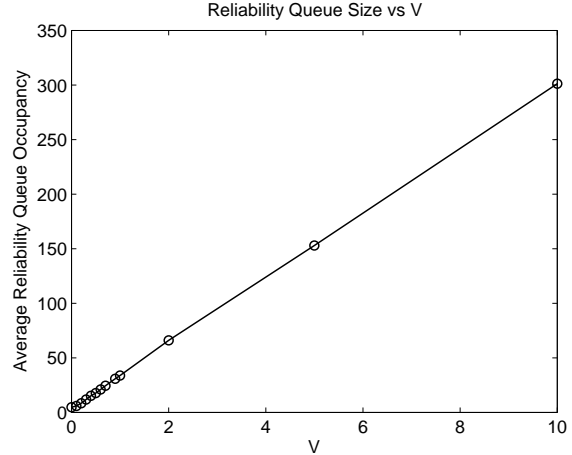}
\caption{Average Reliability Queue Occupancy vs. V.}\label{fig:fig3c}
\end{figure}

\underline{\emph{Existence of an Optimal Stationary, Randomized Policy}}: Assuming
i.i.d. $\mathcal{T}(t)$  states,
%\footnote{A similar statement can be
%made for more general Markov modulated $\mathcal{T}(t)$
%\cite{neely-NOW}. For simplicity, here we consider the i.i.d. case},
there exists a stationary randomized policy $\pi$ that chooses
feasible control action $\mathcal{I}^{\pi}(t)$ and power allocations ${P}_i^{\pi}(t)$ for
all $i \in \mathcal{\widehat{R}}$ every slot purely as a function of the
current channel state $\mathcal{T}(t)$
%and independent of queue backlog
and yields the following for some $\epsilon > 0$:
\begin{align}
&\expect{\Phi_s^{\pi}(t)} \geq \rho_s \lambda_s  + \epsilon  \label{eq:stat0}\\
&\expect{P_i^{\pi}(t)} + \epsilon \leq P_i^{avg}
\label{eq:stat1}\\
&\alpha_s \expect{\Phi_s^{\pi}(t)} - \sum_{i\in \mathcal{N}} \beta_i
\expect{P_i^{\pi}(t)} = \alpha_s r^*_s- \sum_{i\in \mathcal{N}}
\beta_ie^*_i   \label{eq:stat2}
\end{align}
Let $\textbf{\emph{Q}}(t) = (Z_s(t), X_i(t)) \; \forall i \in \mathcal{\widehat{R}}$ 
%\textbf{\emph{Q}}(t) = (Z_s(t), \overrightarrow{X_i}(t))$
represent the collection of these queue backlogs in timeslot $t$. We
define a quadratic Lyapunov function: 
%$L(Q(t)) \defequiv \frac{1}{2}\Big[ Z^2_s(t) + \sum_{i \in \mathcal{\widehat{R}}} X_i^2(t)\Big]$.
\begin{align*}
L(\textbf{\emph{Q}}(t)) \defequiv \frac{1}{2}\Big[ Z^2_s(t) +
\sum_{i \in \mathcal{\widehat{R}}} X_i^2(t)\Big]
\end{align*}

Also define the conditional Lyapunov drift $\Delta(\textbf{\emph{Q}}(t))$ as follows: 
%$\Delta(Q(t))  \defequiv \expect{L(Q(t+1)) - L(Q(t))|Q(t)}$.
\begin{align*}
\Delta(\textbf{\emph{Q}}(t))  \defequiv \expect{L(\textbf{\emph{Q}}(t+1)) - L(\textbf{\emph{Q}}(t))|\textbf{\emph{Q}}(t)}
\end{align*}

Using queueing dynamics (\ref{eq:p1u1}), (\ref{eq:p1x1}), the
Lyapunov drift under any control policy can be computed as follows:
\begin{align}
\Delta(\textbf{\emph{Q}}(t)) \leq \; B &- Z_s(t) \expect{\Phi_s(t) - \rho_s A_s(t) |\textbf{\emph{Q}}(t)}  \nonumber \\
&- \sum_{i\in \mathcal{\widehat{R}}}X_i(t) \expect{P_i^{avg} - P_i(t)|\textbf{\emph{Q}}(t)} 
\label{eq:p1lpdrift2}
\end{align}
where $B = \frac{1 + \lambda_s^2 \rho_s^2 + \sum_{i\in \mathcal{\widehat{R}}} (P_i^{avg})^2 + (P^{max})^2}{2}$. 

For a given control parameter $V \geq 0$, we subtract a ``reward'' metric 
$V\expect{\alpha_s\Phi_s(t) - \sum_{i\in \mathcal{\widehat{R}}} \beta_i P_i(t)|\textbf{\emph{Q}}(t)}$ 
from both sides of the above inequality to get the following:
\begin{align}
\Delta(\textbf{\emph{Q}}(t)) &- V \expect{\alpha_s\Phi_s(t) - \sum_{i\in \mathcal{\widehat{R}}} \beta_i P_i(t)|\textbf{\emph{Q}}(t)} \leq \; B \nonumber \\
&- Z_s(t) \expect{\Phi_s(t) - \rho_s  A_s(t) |\textbf{\emph{Q}}(t)} \nonumber \\
&- \sum_{i\in \mathcal{\widehat{R}}}X_i(t) \expect{P_i^{avg} - P_i(t)|\textbf{\emph{Q}}(t)} \nonumber \\
&- V \expect{ \alpha_s\Phi_s(t) - \sum_{i\in \mathcal{\widehat{R}}} \beta_i P_i(t)|\textbf{\emph{Q}}(t)} 
\label{eq:p1lpdrift3}
\end{align}

From the above, it can be seen that the dynamic control algorithm
(\ref{eq:p1ssp3}) is designed to take a control action that
minimizes the right hand side of (\ref{eq:p1lpdrift3}) over all
possible options every slot, including the stationary policy $\pi$.
Thus, using (\ref{eq:stat0}), (\ref{eq:stat1}), (\ref{eq:stat2}), we
can write the above as:
\begin{align}
\Delta(\textbf{\emph{Q}}(t)) &- V \expect{\alpha_s\Phi_s(t) - \sum_{i\in \mathcal{\widehat{R}}} \beta_i P_i(t)|\textbf{\emph{Q}}(t)} \leq  B \nonumber\\
&- Z_s(t)\epsilon  - \sum_{i\in \mathcal{\widehat{R}}}X_i(t) \epsilon -V \alpha_s r^*_s- \sum_{i\in \mathcal{\widehat{R}}} \beta_ie^*_i
 \label{eq:p1lpdrift4}
\end{align}
Theorem $1$ now follows by a direct application of the Lyapunov
optimization Theorem \cite{neely-NOW}.

\section*{Appendix B -- Solution to Non-Regenerative DF orthogonal using KKT conditions}

We ignore the constant terms in the objective. It is easy to see that the first constraint in
(\ref{eq:nonregen_dfortho2}) must be met with equality. The
Lagrangian is given by:
\begin{align*}
\mathcal{L} = &(X_s + V \beta_s)P_s + \sum_{i\in \mathcal{U}_k} (X_i + V \beta_i) P_i - \lambda_s (P_s - P_s^{\mathcal{U}_k}) 
\\ & - \sum_{i\in \mathcal{U}_k}\lambda_iP_i + \beta_s(P_s - P_s^{max})+ \sum_{i\in \mathcal{U}_k} \beta_i(P_i - P_i^{max})
\\ & + \nu\Big[\log(1+{\theta_sP_s}) + \sum_{i\in \mathcal{U}_k} \log(1+\theta_iP_i) - \frac{mR}{W}\Big]
\end{align*}
where $\theta_s = \frac{m}{W}|h_{sd}|^2, \theta_i =
\frac{m}{W}|h_{id}|^2$. The KKT conditions for all $i\in
\mathcal{U}_{k}$ are:
\begin{align*}
& \lambda_s^*(P_s^* - P_s^{\mathcal{U}_k}) = 0 \qquad \lambda_i^*P_i^* = 0 \\
& \beta_s^*(P_s^* - P_s^{max}) = 0 \qquad \beta_i^* (P_i^* - P_i^{max}) = 0 \\
& \lambda_s^*, \lambda_i^*, \beta_s^*, \beta_i^* \geq 0 \\
& (X_s + V \beta_s) - \lambda_s^* + \beta_s^* + \frac{\nu^* \theta_s}{1 + \theta_s P_s^*} = 0 \\
& (X_i + V \beta_i) - \lambda_i^* + \beta_i^* + \frac{\nu^* \theta_i}{1 + \theta_i P_i^*} = 0
\end{align*}
If $\nu^* > 0$, then we must have that $\lambda_s^* - \beta_s^* > 0$
and $\lambda_i^* - \beta_i^* > 0$ for all $i$. This would mean that
$P_s^* = P_s^{\mathcal{U}_k}$ and $P_i^* = 0$. For some $\nu^* \leq
0$, we have three cases:

\begin{enumerate}
\item If $\lambda_i^* = \beta_i^*$, we get $P_i^* = \frac{-\nu^*}{X_i + V \beta_i} - \frac{1}{\theta_i}$
\item If $\lambda_i^* > \beta_i^*$, then we must have $\lambda_i^* > 0$ and we get $P_i^* = 0$
\item If $\lambda_i^* < \beta_i^*$, then we must have $\beta_i^* > 0$ and we get $P_i^* = P_i^{max}$
\end{enumerate}
Similar results can be obtained for $P_s^*$. Combining these, we get:

$P_s^* = \Big[\frac{-\nu*}{X_s + V \beta_s} - \frac{1}{\theta_s}\Big]_{P_s^{\mathcal{U}_k}}^{P_s^{max}}$ 
$P_i^* = \Big[\frac{-\nu*}{X_i + V \beta_i} - \frac{1}{\theta_i}\Big]_0^{P_i^{max}}$

where $[X]_0^{P_{max}}$ denotes $\min[\max(X, 0), P_{max}]$

\section*{Appendix C -- Solution to AF orthogonal using KKT conditions}

It is easy to see that the first constraint in (\ref{eq:afortho3})
must be met with equality. The Lagrangian is given by:
\begin{align*}
\mathcal{L} = &\sum_{i\in \mathcal{R}_{s}} (X_i + V\beta_i) P_i - \sum_{i\in \mathcal{R}_{s}}\lambda_iP_i 
+ \sum_{\in \mathcal{R}_{s}} \beta_i(P_i - P_i^{max}) \\
& + \nu \Big[\sum_{\in \mathcal{R}_{s}} \frac{P_s^2|h_{si}|^4 + P_s|h_{si}|^2W/m}{|h_{si}|^2P_s + |h_{id}|^2P_i + W/m}- \theta' \Big]
\end{align*}
The KKT conditions for all $i\in \mathcal{R}_{s}$ are:
\begin{align*}
& \lambda_i^* P_i^* = 0 \qquad \beta_i^* (P_i^* - P_i^{max}) = 0 \qquad \lambda_i^*, \beta_i^* \geq 0 \\
& (X_i + V \beta_i) - \lambda_i^* + \beta_i^* = \frac{\nu^*|h_{id}|^2(P_s^2|h_{si}|^4 + P_s|h_{si}|^2 W/m)}{(|h_{si}|^2P_s 
+ |h_{id}|^2P_i^* + W/m)^2} 
\end{align*}
If $\nu^* < 0$, then we must have that $\lambda_i^* - \beta_i^* > 0$
for all $i$. This would mean that $P_i^* = 0$. For some $\nu^* \geq
0$, we have three cases:
\begin{enumerate}
\item If $\lambda_i^* = \beta_i^*$, we get $P_i^* = \sqrt{\frac{\nu^*(P_s^2|h_{si}|^4 + P_s|h_{si}|^2 W/m)}{(X_i + V \beta_i)|h_{id}|^2}} - 
\frac{P_s|h_{si}|^2 + W/m}{|h_{id}|^2}$
\item If $\lambda_i^* > \beta_i^*$, then we must have $\lambda_i^* > 0$ and we get $P_i^* = 0$
\item If $\lambda_i^* < \beta_i^*$, then we must have $\beta_i^* > 0$ and we get $P_i^* = P_i^{max}$
\end{enumerate}

Combining these, we get:

$P_i^* = \Big[\sqrt{\frac{\nu^*(P_s^2|h_{si}|^4 + P_s|h_{si}|^2W/m)}{(X_i + V \beta_i)|h_{id}|^2}} 
- \frac{P_s|h_{si}|^2 + W/m}{|h_{id}|^2}\Big]_0^{P_i^{max}}$ 
where $[X]_0^{P_{max}}$ denotes $\min[\max(X, 0), P_{max}]$

\end{document}